# Unraveling persistent urban-rural gaps: A long-term provincial analysis of residential heating and cooling loads

Qinwen Tang [1], Ran Yan [2], Nan Zhou [3], Minda Ma [1*, 4]

1. School of Architecture and Urban Planning, Chongqing University, Chongqing, 400045, PR China
2. School of Management Science and Real Estate, Chongqing University, Chongqing, 400045, PR China
3. Energy and Resources Group, University of California, Berkeley, CA 94720, United States
4. Lawrence Berkeley National Laboratory, Berkeley, CA 94720, United States

- Corresponding author: Dr. Minda Ma, Email: maminda@lbl.gov; minda.ma@cqu.edu.cn
  Homepage: https://globe2060.org/MindaMa/

- Preprint of this submission is available at: https://arxiv.org/abs/2512.16779

## Graphical abstract

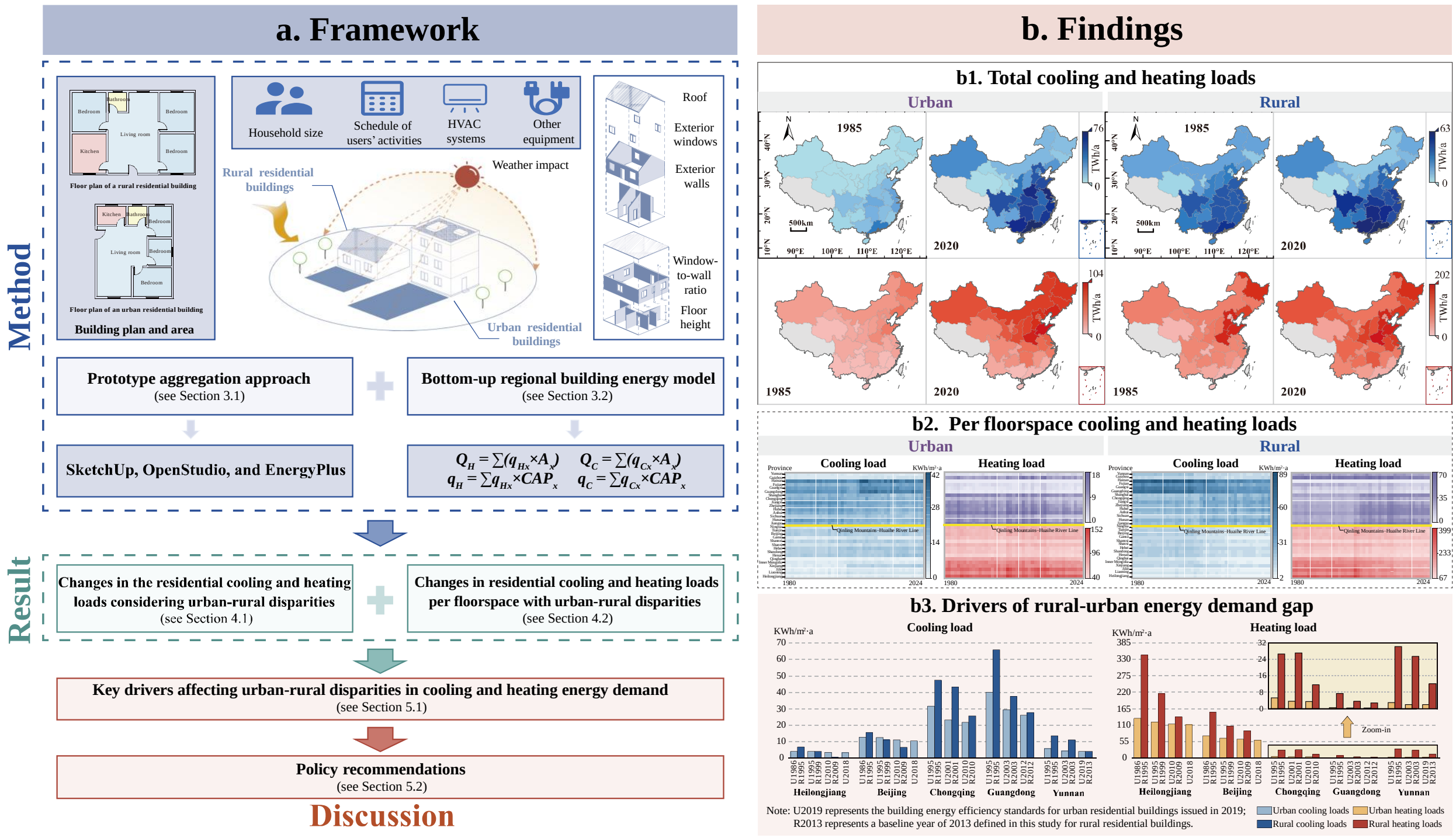

**Graphical Abstract:** A long-term provincial analysis unraveling persistent urban-rural gaps in residential space heating and cooling loads. The graph shows (a) the study framework; (b1) total space cooling and heating loads of urban and rural residential buildings by province in 1985 and 2020; (b2) loads per floorspace in urban and rural China (1980-2024); and (b3) decomposition of cooling and heating loads of urban and rural residential buildings by construction period (observed in 2024).

## Highlights

- Proposed a bottom-up model to assess residential heating and cooling loads across 30 provinces.
- Urban-rural total load disparities narrowed, but rural load intensity remained 20%-100% higher.
- Rural heating loads declined (-0.7%/a), while urban cooling loads increased (0.4%/a) in 1980-2024.
- Building stock shifted as urban surpassed rural in 22 provinces, reshaping national load patterns.
- Envelope-performance gaps are a major driver of persistent urban-rural load-intensity disparities.

## Abstract

Climate change and rising thermal comfort demand make residential heating and cooling central to building-sector decarbonization. This study presents a bottom-up modeling framework to estimate residential heating and cooling loads across 30 Chinese provinces. The model, developed using EnergyPlus simulations of representative building prototypes, captures heating and cooling load patterns in both urban and rural housing over the period 1980–2024. The results indicate that: (1) In 2020, Guangdong recorded the highest cooling loads (76.5 TWh/a urban; 63.0 TWh/a rural). Henan exhibited the highest rural heating load (174.6 TWh/a), while urban heating loads were highest in Liaoning and Shandong. (2) Between 1980 and 2024, average urban cooling loads increased from 12.4 to 15.1 kWh/m$^2$·a, whereas rural cooling loads declined from 22.63 to 19.87 kWh/m$^2$·a. Urban heating loads decreased from 44.08 to 39.92 kWh/m$^2$·a, and rural heating loads declined more markedly from 100.15 to 72.42 kWh/m$^2$·a. (3) Urban residential floor area has exceeded rural stock in 22 provinces in recent years, compared with only four provinces in 2000. Moreover, the existence of 12 urban energy-efficiency standards versus a single rural standard reflects a persistent regulatory disparity. Counterfactual and factor-contribution analyses further indicate that the associated envelope-performance gap is a major contributor to urban-rural load-intensity differences, while building form has a more context-dependent influence. The proposed framework provides a replicable basis for region-specific clean heating strategies and differentiated building standards to support carbon neutrality.

## Abbreviations notation

| | |
|---|---|
| C | Cold zone |
| HSCW | Hot summer and cold winter zone |
| HVAC | Heating, ventilation, and air conditioning |
| HSWW | Hot summer and warm winter zone |
| $kWh/m^2/a$ | Kilowatt-hours per square meter per year |
| M | Mild zone |
| $m^2$ | Square meters |
| SC | Severe cold zone |
| TWh/a | Terawatt-hours per year |
| $W/m^2$ | Watts per square meter |

**Nomenclature**

| | |
|---|---|
| $A_x$ | The total floorspace of residential buildings constructed in period x |
| $CAP_x$ | The proportion of residential building floorspace constructed in year x |
| O.S | Occupied with occupants asleep |
| O.W | Occupied with occupants awake |
| $Q_C$ | The total regional cooling loads of residential buildings |
| $q_C$ | The average cooling loads of residential buildings |
| $q_{Cx}$ | The cooling loads of buildings constructed in year x |
| $Q_H$ | The total regional heating loads of residential buildings |
| $q_H$ | The average heating loads of residential buildings |
| $q_{Hx}$ | The heating loads of buildings constructed in year x |
| U | Unmanned |

## 1. Introduction

The building sector has long been a major contributor to national final energy use and carbon emissions [1, 2]. Studies show that buildings account for 36% of global energy consumption, while the building sector contributes nearly 40% of total global carbon emissions [3, 4]. Much of this impact is attributable to residential buildings, which constitute the largest share of the global building stock [5]. As economies grow and urbanization accelerates, rising household incomes have driven a surge in demand for indoor thermal comfort. This shift has rapidly transformed heating and cooling from luxury amenities into basic necessities [6]. In 2023, space heating was the primary household energy use in the European Union, accounting for 62.5% of final energy consumption in the residential sector [7]. Moreover, global warming and the increasing frequency of extreme weather events have expanded cooling demand across multiple climate zones. By 2100, global energy demand for cooling is projected to increase by 220% [8].

In recent years, scholars across different countries have conducted extensive research on building heating and cooling loads [9, 10], including simulations of heating demand in northern China [11] and assessments of climate change impacts on cooling loads [12]. These studies have provided important methodological foundations and empirical evidence for analyzing building thermal loads. **However**, existing research remains limited in scope: most studies focus on typical urban samples or a single urban-rural category, which constrains the ability to examine long-term urban-rural disparities in heating and cooling loads at the provincial level. To address these gaps, this study pioneers the reconstruction of China's residential heating and cooling loads across both provincial and urban-rural dimensions from 1980 to 2024. Specifically, the study addresses three key scientific questions:

- How have residential heating and cooling loads evolved across China over time?
- What urban-rural differences exist in building thermal loads, and what are their key driving factors?
- How can the findings inform future building energy conservation policies for urban and rural areas?

To address these research gaps, this study employs a prototype building approach to construct urban and rural residential building models for different construction periods in each province. Combined with a bottom-up calculation method, the heating and cooling loads of residential buildings in 30 Chinese provinces from 1980 to 2024 are quantified. Through a detailed analysis of spatiotemporal differences in urban-rural loads, the underlying driving mechanisms—including socioeconomic factors, building characteristics, and climate adaptation behaviors—are identified, providing a scientific basis for formulating differentiated energy conservation policies for residential buildings.

**The primary contribution of this study** is the development of a multifactor residential building prototype system that integrates urban-rural differences, building construction periods, and regional characteristics within an EnergyPlus-based simulation framework. This framework enables the quantification of spatiotemporal variations in the thermal loads of urban and rural residential buildings at the provincial scale. By adopting a bottom-up approach and explicitly examining the underlying driving mechanisms, the study provides a robust scientific basis for formulating province-specific building energy-saving policies.

The remainder of this paper is organized as follows. Section 2 reviews the relevant literature. Section 3 describes the methodological framework and data sources. Section 4 presents the temporal and spatial trends in urban and rural heating and cooling loads across China's 30 provinces. Section 5 further analyzes these loads by building prototype and examines the key drivers underlying their divergent urban-rural trends. Finally, Section 6 summarizes the main conclusions of the study.

## 2. Literature review

The mainstream tools for simulating total building loads at the regional scale adopt two distinct approaches: top-down or bottom-up. In macroscale analyses, top-down methods in the building energy field utilize aggregated data at the macro level, including regional energy statistics and economic indicators, to analyze energy consumption [13]. This approach offers high computational efficiency and is suitable for trend analysis and forecasting. For example, Duanmu, et al. [14] adopted a top-down method to rapidly predict hourly regional cooling loads at the urban energy planning stage on the basis of macrolevel building information. Hu, et al. [15] proposed a top-down, data-driven method that uses a segmented multiple regression model to decompose building thermal loads from grid electricity data, thereby predicting heating and cooling electricity demand. Although top-down methods can quickly produce regional load forecasts from aggregated data, the data aggregation process inevitably masks important details, including differences in building construction periods and spatial characteristics, the impact of interregional climatic variations, and the spatiotemporal variability of occupant behavior patterns. This limits their applicability in scenarios requiring detailed analysis.

For more detailed load prediction, particularly in the evaluation of various building characteristics or energy-saving measures, the bottom-up approach provides significant advantages. By integrating detailed building-level data into the modeling process, the bottom-up approach can produce more accurate spatial distributions of heating and cooling loads for regional building stocks and better reflect interbuilding differences, despite its greater computational complexity. For example, Li, et al. [16] proposed a bottom-up engineering method to establish residential building energy models for predicting heating and cooling loads in China's hot summer and cold winter zones under current and future conditions, thereby providing guidance for building design and retrofit policies. Kim, et al. [17] developed a novel physics-based bottom-up modeling framework for commercial building electricity loads. This framework accounts for large population dwelling profiles and heterogeneity in building systems, enabling simulation of time series electricity load curves for the commercial sector to support policy formulation. Zi, et al. [18] adopted a bottom-up approach that integrates geospatial data to analyze building characteristics, perform energy simulations of cooling loads and solar collection, and quantify performance and decision-making to achieve sustainable urban energy systems. Therefore, to more accurately calculate regional building heating and cooling loads and analyze their influencing factors, this study adopts a bottom-up approach.

Calculating building heating and cooling loads is typically achieved through dynamic simulation methods. These methods generally rely on specialized simulation software for numerical computation and system analysis. These methods consider multiple influencing factors, including climate, building systems, and occupant behavior, to simulate their interactions and accurately compute building loads [19]. Commonly used simulation tools include EnergyPlus, the Integrated Environmental Solutions Virtual Environment, DesignBuilder and the Dynamic Energy Simulation Tool. In response to the needs of dynamic load simulation, EnergyPlus demonstrates a notable accuracy advantage among its counterparts through rigorous solutions of heat transfer equations and coupled system simulations. Li, et al. [20] employed EnergyPlus as the core engine of the Urban Modeling Interface tool to perform physical modeling of heating and cooling energy consumption for 442 buildings in Chongqing. High-precision data were generated to train machine learning models, ultimately enabling rapid and accurate prediction of energy consumption at the building cluster level and assessment of energy-saving potential. Yang, et al. [21] adopted EnergyPlus as the core simulation tool, incorporating three years of hourly weather data from 15 cities in eastern China to physically model the cooling energy use of typical residential buildings, analyze the impacts of heatwaves and urban microclimates on cooling demand, and reveal the relationships between temperature, humidity, and sensible and latent cooling loads, thereby providing a basis for assessing building cooling energy consumption. Guo, et al. [22] utilized EnergyPlus as the core engine of the white-box building model, coupled with a gray-box model in MATLAB, to enable co-simulation for integrating building design optimization and predictive control, with life-cycle cost analysis providing accurate data support for the energy-efficient optimization of building envelopes. Razzano, et al. [23] employed EnergyPlus within a high-fidelity co-simulation environment to model the building envelope and loads, providing accurate data support for performance comparisons between deep reinforcement learning and rule-extraction-based control methods. Research has demonstrated that EnergyPlus provides accurate and widely applicable support for building energy modeling across diverse contexts. Therefore, this study, which is based on EnergyPlus, combines a bottom-up approach with classification and summary according to the typical characteristics of buildings [24, 25], develops representative physical models, and extrapolates the simulation results of these typical buildings to the regional scale [26, 27].

China covers a vast territory, with substantial differences in resource endowment and stages of development among provinces. This spatial heterogeneity lends unique significance to research conducted at the provincial scale. However, with respect to residential building heating and cooling loads, most existing studies focus on either urban or rural residential buildings in selected representative provinces [28-31]. Few studies have conducted long-term, nationwide comparisons of urban and rural loads at the provincial scale. To address this gap, the present study develops a bottom-up framework based on EnergyPlus and a prototype-building approach to simulate and quantify the heating and cooling loads of urban and rural residential buildings across 30 Chinese provinces from 1980 to 2024, thereby enabling a detailed comparison of urban-rural load differences.

**The principal contributions of this work are summarized as follows:**

- **This study conducts dynamic simulations and comparative analyses of heating and cooling loads in urban and rural residential buildings across China during their operational phase (1980-2024).** Based on a bottom-up analytical framework, it employs a building prototype approach implemented through the EnergyPlus simulation platform. By integrating multidimensional data—including climate zone characteristics, building thermal performance, and energy-use behavior patterns—the study establishes a robust technical pathway for accurate carbon accounting of urban and rural residential buildings across different climate zones. These results provide a scientific assessment basis to support the formulation of differentiated emission-reduction policies and offer methodological support for precise carbon emission quantification in both urban and rural buildings.
- **This study presents a provincial-scale comparative analysis of heating and cooling loads in urban and rural residential buildings and establishes an efficient upscaling methodology from representative building prototypes to provincial-level aggregates.** It quantifies the long-term evolution of thermal loads in China's urban and rural housing stock over a 44-year period and reveals dynamic divergence patterns in load characteristics between urban and rural buildings across different climate zones. Furthermore, these long-term provincial-scale heating and cooling load trajectories provide critical theoretical foundations and practical guidance for regionally tailored building energy-efficiency policy formulation.

# 3. Materials and methods

To calculate the urban and rural cooling and heating loads at the provincial level over the period 1980-2024, a bottom-up regional building thermal load model based on the prototype building approach was employed for this analysis in this section.

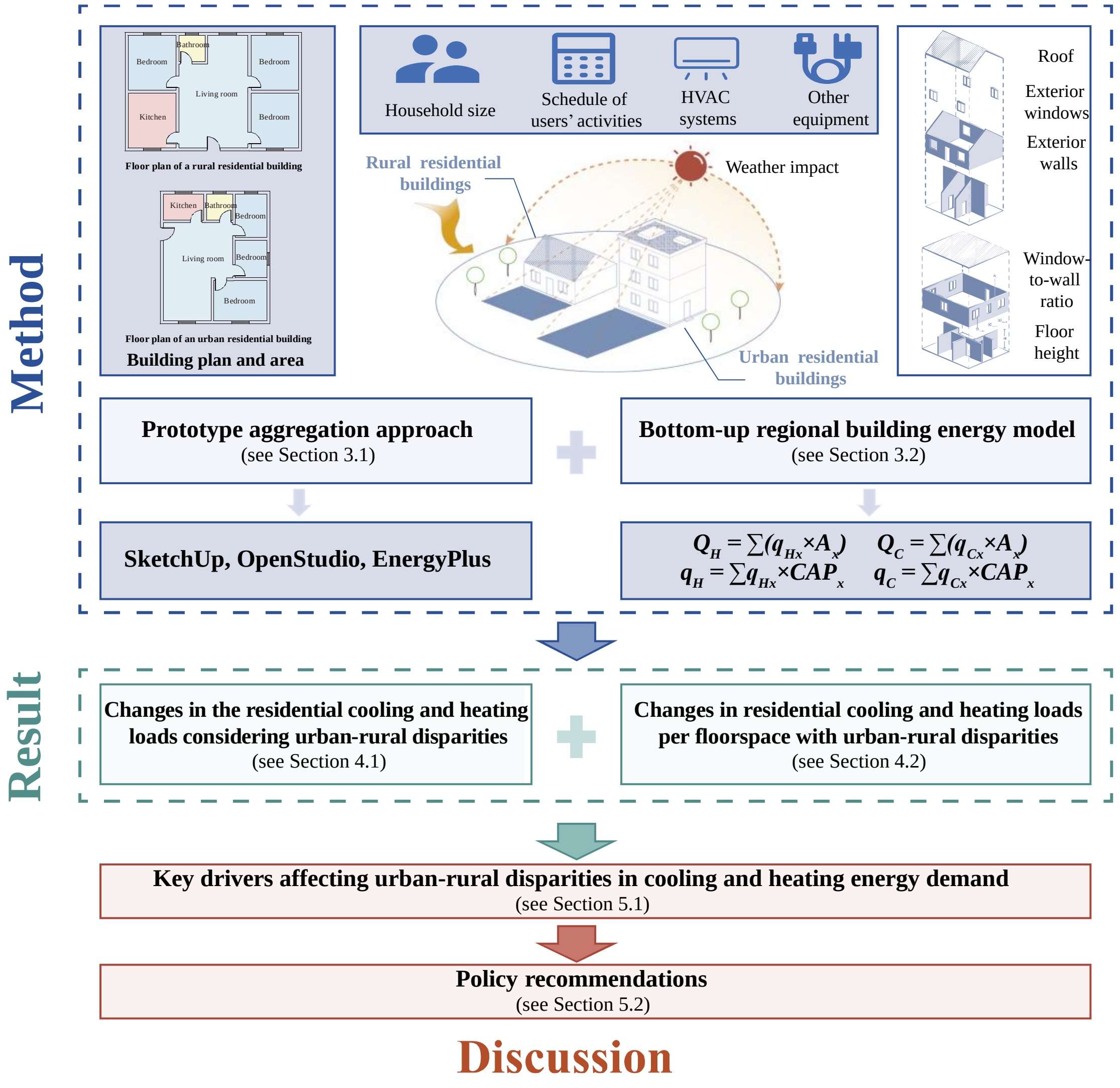

**Fig. 1.** Framework of this study.

## *3.1. Prototype aggregation approach*

The prototype building method was used to define representative structures that encapsulated the architectural characteristics of the study area, followed by their analysis and computational evaluation. The use of prototype buildings in prototype-based modeling is a widely adopted approach for regional building thermal load analysis [32, 33]. The heating and cooling loads of a building are influenced primarily by its physical form, internal and external heat gains, and the Heating, Ventilation, and Air Conditioning (HVAC) system [34-37]. Therefore, on the basis of these influencing factors, the construction of the original building model can be summarized into three parts: actual building physical form, building envelope parameters, and occupancy activity prototypes.

The physical form and envelope of a building are typically related to its climatic zone. To simplify the prototype building samples, China’s building climate zones, which includes the severe cold zone (SC), cold zone (C), hot summer and cold winter zone (HSCW), hot summer and warm winter zone (HSWW), and mild zone (M), was adopted as the basis for constructing building prototypes. Provinces within the same climatic zone were represented by the same building prototype. Given that most residential buildings in China are rectangular [38], the floor plans of dwellings in both urban and rural areas were assumed to be rectangular in this study. The construction of the floor plan layouts was based on authentic layouts obtained from the residential building real estate market and literature review while also considering the privacy requirements and convenience of the occupants.

Urban residential buildings are typically multistory or high-rise multifamily dwellings [28]. The selected typical household was located on the middle floor of the building, featuring three exterior walls and one interior wall. The entrance door was placed on the interior wall, while exterior windows were installed on the south-facing, east-facing, and north-facing exterior walls (as illustrated in Fig. 2). The typical rural residential building was a detached, self-built structure with exterior walls on all four sides. The external windows were typically installed on south-facing and north-facing walls, whereas the east and west exterior walls were generally windowless (as illustrated in Fig. 2). The roof structure was typically a pitched roof design.

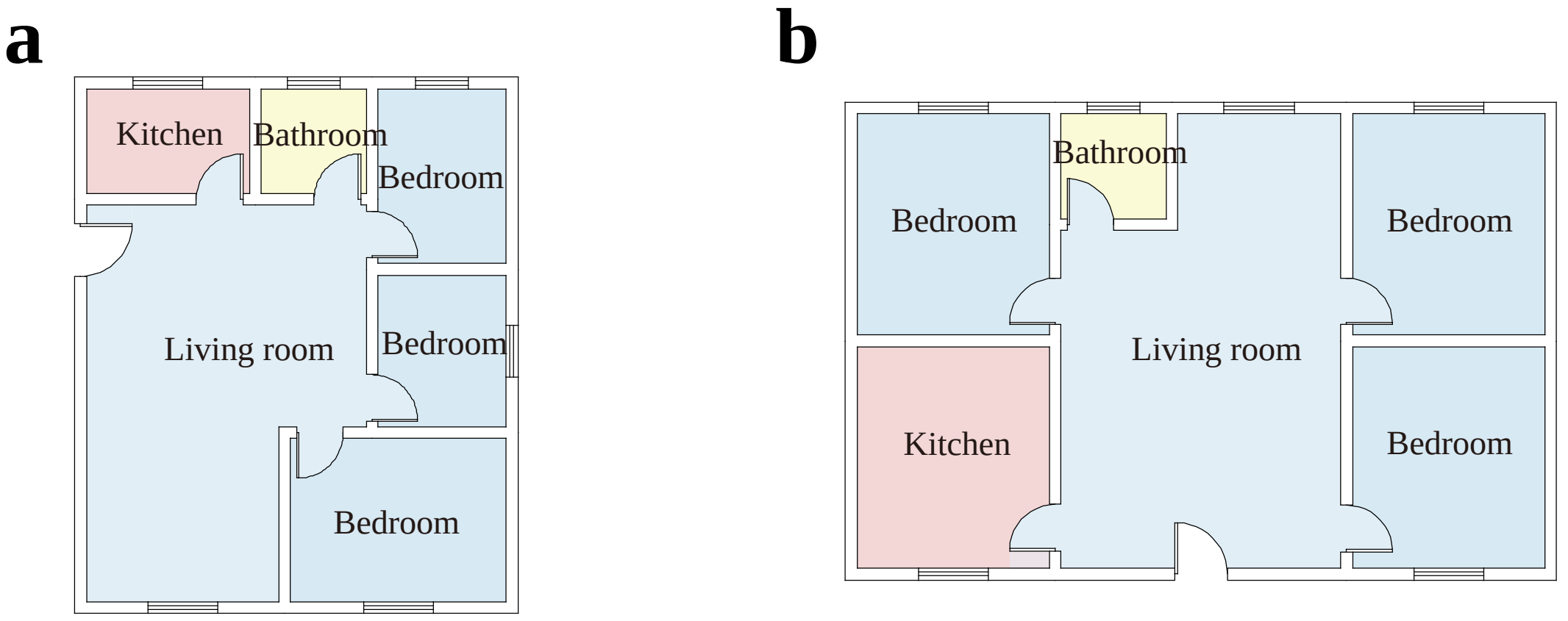

**Floor plan of an urban residential building** **Floor plan of a rural residential building**

**Fig. 2.** (a) Urban residential building plan and (b) rural residential building plan.

The architectural form is closely related to the building's construction period [39]. Given the broad research time span, building prototype data were obtained from building energy efficiency indicators across the five distinct climate zones. These were divided into two periods, which were demarcated roughly around the year 2000, to facilitate subsequent research. By collecting the floorspace, building aspect ratio, building floor height, and window-to-wall ratio (see Tables B1 and B2 of Appendix B), physical models of urban and rural residential buildings in five thermal divisions and in two construction periods were developed via SketchUp and OpenStudio, respectively. For urban residential buildings, the building envelope parameters of structures built in different eras were determined by the respective regional building energy efficiency design standards. Therefore, the classification of residential building construction periods was based on updates of energy-saving design standards for residential buildings in each climate zone [40-43]. The specific classifications are as follows:

The construction periods for residential buildings in the zones of SC and C were categorized into 1980-1995 (including buildings completed in 1995), 1996-2010 (including buildings completed in 2010), 2011-2018 (including buildings completed in 2018), and 2019-2024 (including buildings completed in 2024). These periods correspond to the first edition of the energy-saving design standard JGJ 26-1986, the second edition JGJ 26-1995, the third edition JGJ 26-2010, and the fourth edition JGJ 26-2018. For residential buildings in zones characterized by the HSCW, the construction periods were categorized into 1980-2001 (including buildings completed in 2001), 2002-2010 (including buildings completed in 2010), and 2011-2024 (including buildings completed in 2024).

These periods correspond to the first edition of the energy-saving design standard JGJ 134-2001 and the second edition JGJ 134-2010, respectively. For zones characterized by the HSWW, the construction periods were divided into 1980-2003 (encompassing residential buildings constructed up to and including 2003), 2004-2012 (encompassing residential buildings constructed up to and including 2012), and 2013-2024 (encompassing residential buildings constructed up to and including 2024). These periods correspond to the first edition of the energy-saving design standard JGJ 75-2003 and the second edition JGJ 75-2012, respectively. The M zone was treated as a special case, corresponding only to the energy-saving design standard JGJ 475-2019, and was divided with reference to the HSWW zone into 1980-2003, 2004-2019, and 2020-2024, in line with the same energy-saving design standard.

**Table 1.** Development and thermal performance of urban residential buildings.

| Thermal climate zones for buildings | Year of construction | Exterior wall heat transfer coefficient W/($m^2$·K) | Exterior window heat transfer coefficient W/($m^2$·K) | Solar heat gain coefficient of exterior windows | Number of air changes (times/h) |
|---|---|---|---|---|---|
| Severe cold zones (Severe cold zones A, B, and C) | 1980-1995 | 0.90-0.63 | 3.26-1.60 | 0.60 | 1 |
| | 1996-2010 | 0.68-0.52 | 4.00-2.50 | 0.60 | 1 |
| | 2011-2018 | 0.30 | 1.60 | 0.50 | 0.5 |
| | 2019-2024 | 0.25 | 1.40 | 0.45 | 0.5 |
| Cold zones (Cold zones A and B) | 1980-1995 | 1.10-0.90 | 3.26-2.80 | 0.60 | 1 |
| | 1996-2010 | 0.90-0.68 | 3.26-2.80 | 0.60 | 1 |
| | 2011-2018 | 0.60 | 2.50 | 0.50 | 0.5 |
| | 2019-2024 | 0.45 | 2.00 | 0.45 | 0.5 |
| Hot summer and cold winter zones | 1980-2001 | 1.97 | 5.78 | 0.85 | 2 |
| | 2002-2010 | 1.03 | 2.80 | 0.48 | 1 |
| | 2011-2024 | 0.83 | 2.67 | 0.34 | 1 |
| Hot summer and warm winter zones | 1980-2003 | 2.50 | 5.90 | 0.60 | 1.5 |
| | 2004-2012 | 2.00 | 3.00 | 0.50 | 1 |
| | 2013-2024 | 1.50 | 2.50 | 0.45 | 1 |
| Mild zone | 1980-2003 | 1.31 | 5.78 | 0.60 | 1.5 |
| | 2004-2019 | 1.20 | 3.20 | 0.50 | 1 |
| | 2020-2024 | 0.80 | 2.50 | 0.45 | 1 |

In setting the parameters for the exterior envelope of urban residential buildings, it was assumed that all interior surfaces (including interior walls, floor slabs, and ceilings) were adiabatic to simplify the simulation process. This scenario represents a situation in which the heating and cooling regulation behavior of neighboring domestic households was similar to the preferences of the occupants, and the room temperatures between households were very close to each other. The thermal performance of the building envelope for urban residential buildings of different construction periods (as shown in Table 1) was determined in accordance with the relevant specifications to ensure the accuracy and representativeness of the building prototypes. In rural China, residential buildings were predominantly self-built, especially before 2000. Owing to economic underdevelopment, thermal insulation measures were rarely adopted in building envelopes during this period. Energy efficiency standards for rural residential buildings in China were not formally established until 2013 [44]. Considering the realities of rural areas and relevant building codes, time-based segmentation of the building envelope for different climate zones was proposed (as shown in Table 2).

**Table 2.** Development and thermal performance of rural residential buildings.

| Thermal climate zones for buildings | Year of construction | Exterior wall heat transfer coefficient W/($m^2$·K) | Roof heat transfer coefficient W/($m^2$·K) | Exterior window heat transfer coefficient W/($m^2$·K) | Solar heat gain coefficient of exterior windows | Number of air change (times/h ) |
|---|---|---|---|---|---|---|
| Severe cold zones (Severe cold zones A, B, and C) | 1980-1999 | 1.54 | 2.39 | 4.70 | 0.60 | 1 |
| | 2000-2009 | 0.96 | 0.96 | 4.70 | 0.60 | 1 |
| | 2010-2024 | 0.45 | 0.35 | 2.00 | 0.45 | 0.5 |
| Cold zones (Cold zones A and B) | 1980-1999 | 1.54 | 2.39 | 4.70 | 0.60 | 1 |
| | 2000-2009 | 0.96 | 0.96 | 4.70 | 0.60 | 1 |
| | 2010-2024 | 0.60 | 0.50 | 2.70 | 0.45 | 0.5 |
| Hot summer and cold winter zones | 1980-2001 | 2.07 | 4.04 | 4.70 | 0.85 | 2 |
| | 2002-2010 | 2.03 | 3.14 | 4.70 | 0.48 | 1 |
| | 2011-2024 | 1.80 | 0.94 | 3.20 | 0.34 | 1 |
| Hot summer and warm winter zones | 1980-2003 | 2.56 | 2.62 | 6.40 | 0.60 | 1.5 |
| | 2004-2012 | 1.50 | 1.00 | 6.40 | 0.60 | 1 |
| | 2013-2024 | 1.20 | 0.80 | 4.00 | 0.50 | 1 |
| Mild zone | 1980-2003 | 2.07 | 4.04 | 6.40 | 0.60 | 1.5 |

| Thermal climate zones for buildings | Year of construction | Exterior wall heat transfer coefficient W/(m$^2$·K) | Roof heat transfer coefficient W/(m$^2$·K) | Exterior window heat transfer coefficient W/(m$^2$·K) | Solar heat gain coefficient of exterior windows | Number of air change (times/h) |
|---|---|---|---|---|---|---|
| | 2004-2013 | 2.03 | 3.14 | 4.70 | 0.50 | 1 |
| | 2014-2024 | 1.80 | 1.00 | 3.20 | 0.45 | 1 |

The thermal conditions within a building are closely related to the operation of the HVAC system and occupancy patterns [45]. Occupants who worked or attended school were typically at home during non-working or non-school hours, whereas retired occupants spent more time at home. In China, the most common heating and cooling usage pattern is part-time and partial-space (i.e., only heating or cooling the spaces in use), so the duration of occupancy significantly impacts cooling and heating thermal load. Personnel activities directly affect the cooling and heating loads inside a building [46]; therefore, occupant use schedules are required for building cooling and heating load simulations. Considering whether the occupants were working/studying or not, the occupant timetable was set to account for both long-term home and non-long-term home occupants [16]. The operation of the equipment was considered when the indoor occupants were present and was reduced when the occupants were in a sleeping state. Artificial lighting was required to be turned on after 17:00 [47]. The internal heat sources within the building were set as follows: the artificial lighting density was set to 6 watts per square meter (W/m$^2$), and the equipment density was set to 4.3 W/m$^2$.

**Table 3.** Correspondence between different room types and occupant schedules.

| Room type | User behavior | 0：00-8：00 | 8：00-12：00 | 12：00-14：00 | 14：00-17：00 | 17：00-22：00 | 22：00-24：00 |
|---|---|---|---|---|---|---|---|
| Event space | At home for long periods of time | O.S | O.W | U | O.W | O.W | U |
| | Not at home for long periods of time | O.S | U | U | U | O.W | U |
| bedchamber | At home for | U | U | O.S | U | U | O.W |

| | | | | | | |
|---|---|---|---|---|---|---|
| long periods of time | | | | | | |
| Not at home for long periods of time | U | U | O.S | U | U | O.W |

Note: Unmanned (U), occupied with occupants awake (O.W.), and occupied with occupants asleep (O.S.). The occupancy period divisions in this table were adapted from [47], with the original time intervals (8:00-12:30 and 12:30-14:00) modified to 8:00-12:00 and 12:00-14:00, respectively.

Given the limited occupancy time in auxiliary function rooms (including kitchens, bathrooms, and storage areas) and the general absence of air conditioning to regulate their temperature, the simulation calculations considered only cooling and heating in bedrooms and activity areas (including living rooms and studies). Furthermore, the HVAC systems in bedrooms and activity areas were activated only during periods of occupancy. Considering China's economic conditions prior to 2000, the HVAC system configuration before 2000 typically included one or two fewer bedrooms or activity areas than did the post-2000 configurations. The activation of the HVAC system was related to human comfort, with a comfortable temperature range of 18-26 °C. Cooling was activated when the summer temperature exceeded 26 °C, and heating was activated when the winter temperature fell below 18 °C [40-43]. In rural areas of northern China and in the HSCW zone south of the Qinling-Huaihe Line, decentralized heating systems are widely used in winter, differing significantly from the centralized heating mode in northern urban areas. In the zones of SC and C, centralized heating was adopted; therefore, residential activities were not considered when the heating schedule was set, and the system was assumed to operate continuously for 24 hours. In addition, both the bedrooms and living areas were equipped with radiators regardless of the building construction period. Accordingly, the indoor heating temperature in zones of SC and C was set at 18-22 °C [40]. These differences are summarized in the appendix (see Table C1 of Appendix C).

### *3.2. Bottom-up regional building energy model*

Section 3.1 introduces the prototype building method. The model constructed via the prototype building approach can represent only buildings constructed during a specific period within the research area. In reality, buildings from various construction periods often coexist in the same space. To study the specific cooling and heating loads in 30 provinces and rural areas of China from 1980-2024, a bottom-up regional building thermal load model was employed. This model uses the calculation results of the prototype buildings and their distribution within the research area to determine the average regional thermal load. EnergyPlus, an advanced building energy simulation software developed by the U.S. Department of Energy, is included in the International Building Performance Simulation Association list of building energy simulation tools. The simulation results of EnergyPlus have been validated through a series of analytical tests, comparative tests, and release and executability tests. EnergyPlus has been widely used in research related to building energy. Therefore, in this study, EnergyPlus was utilized to simulate the heating and cooling loads of various typical prototype buildings. Simulated data were then used to develop a bottom-up regional thermal load model for residential buildings, enabling analysis of the specific cooling and heating loads in urban and rural areas across 30 provinces in China from 1980 to 2024 (see Appendix D for validation against existing studies).

In the simulation, the capital city of each province was selected to represent the typical meteorological conditions (see Appendix E for the weather-data sensitivity analysis). Climate data from 1980-2024 for each province were collected, and the annual specific cooling and heating loads for urban and rural residential buildings were simulated for prototype buildings from different construction periods. Furthermore, considering the thermal load differences among residential buildings from different construction periods and the influence of building prototypes on heating and cooling loads, the division of residential building construction periods was based on updates to the energy-saving design standards for residential buildings in each climate zone. The proportion of residential buildings from different construction periods in the urban and rural areas of each province was calculated annually (see Appendix F for the key-assumption sensitivity analysis).

The total heating and cooling loads of urban and rural residential buildings in each province were calculated via the following equations:

$$Q_H = \sum(q_{Hx} \times A_x) \tag{1}$$

$$Q_C = \sum(q_{Cx} \times A_x) \tag{2}$$

where $Q_H$ and $Q_C$ represent the total regional heating and cooling loads of residential buildings; $q_{Hx}$ and $q_{Cx}$ denote the heating and cooling loads of buildings constructed in year x; and $A_x$ denotes the total floorspace of residential buildings constructed in period x.

The average heating and cooling loads densities for urban and rural residential buildings in each province were calculated as follows:

$$q_H = \sum q_{Hx} \times CAP_x \tag{3}$$

$$q_C = \sum q_{Cx} \times CAP_x \tag{4}$$

where $q_H$ and $q_C$ represent the average heating and cooling loads of residential buildings; $q_{Hx}$ and $q_{Cx}$ denote the heating and cooling loads of buildings constructed in year x; and $CAP_x$ signifies the proportion of residential building floorspace constructed in year x.

### *3.3. Data collection*

Data on the average floorspace per household in urban and rural areas across various provinces of China were obtained from statistical yearbooks published by the National Bureau of Statistics, supplemented with province-specific data. Physical characteristic data, including building aspect ratios, window-to-wall ratios, and floor heights, were collected through onsite surveys and publicly available platforms, such as Zhulong.com. Heat transfer coefficient data for exterior windows, walls, and roofs were extracted from building energy-saving design standards applicable to different thermal-climatic zones. For buildings constructed in different eras, the building envelope parameters were adjusted in accordance with the corresponding energy-saving standards. Post-2000 climate data were obtained from Meteonorm 8 EnergyPlus Weather files, whereas pre-2000 climate data were sourced from the transition from the ECMWF Reanalysis v5 (ERA5) reanalysis dataset (see Appendix G for the continuity check). These datasets included parameters such as temperature, humidity, and solar radiation, which served as input conditions for the thermal load simulations. Internal heat gain data, including occupant activity patterns and equipment usage, were established through literature reviews and actual measurements.

## 4. Results

### *4.1. Changes in the residential cooling and heating loads considering urban-rural disparities*

Fig. 3 presents the spatial distribution of total urban cooling and heating loads across China's 30 provinces for four representative years, including 1985, 2000, 2010, and 2020, with subplots (a) and (b) showing the provincial-level total cooling loads and total heating loads, respectively. Fig. 3a reveals significant spatial heterogeneity and a consistent temporal evolution in the total cooling load of urban residential buildings across Chinese provinces from 1985 to 2020. The spatial distribution clearly decreased from northwest to southeast China. In 1985, only five provinces, including Guangdong, Jiangsu, Zhejiang, Hubei and Hunan, had total cooling loads exceeding 3 Terawatt-hours per year (TWh/a), with Guangdong ranking highest at 6.4 TWh/a, whereas Qinghai recorded the lowest value at only 0.01 TWh/a. Half of the provinces had cooling loads below 1 TWh/a. This regional disparity continued to widen over time: By 2000, the gap between the highest (Guangdong, 21.2 TWh/a) and the lowest (Qinghai, 0.11 TWh/a) reached 21.1 TWh/a. By 2010, the difference between Guangdong (42.5 TWh/a) and Qinghai (0.13 TWh/a) further increased to 42.4 TWh/a. By 2020, Guangdong (76.5 TWh/a) far surpassed second-ranked Jiangsu (32.7 TWh/a), with its lead over Qinghai (0.17 TWh/a) expanding to 76.3 TWh/a. The number of high-load provinces (≥1 TWh/a) rose from 5 in 1985 to 14 in 2020, whereas the number of low-load provinces (<1 TWh/a) decreased from 15 in 1985 to only two (Gansu and Qinghai) by 2020.

Temporally, the total cooling load of urban residential buildings in China exhibited an accelerating growth trend. During the period from 1985 to 2000, all provinces experienced notable annual increases, with Guangdong (+14.8 TWh/a) and Jiangsu (+11.1 TWh/a) leading the nation, whereas Qinghai showed the smallest increase (<0.1 TWh/a). The period 2000-2010 marked a phase of divergence: provinces such as Guangdong (+21.3 TWh/a) and Guangxi (+9.1 TWh/a) maintained rapid growth, whereas Gansu and Liaoning experienced negative growth. Between 2010 and 2020, the cooling load increased dramatically. Five provinces recorded annual growth exceeding 10 TWh/a, notably, Guangdong (+34 TWh/a) and Henan (+15.4 TWh/a), while Heilongjiang was the only province that experienced a decrease. This spatiotemporal evolution reflects the expansion of cooling load from southeastern coastal zones inland, alongside a persistent intensification of regional disparities.

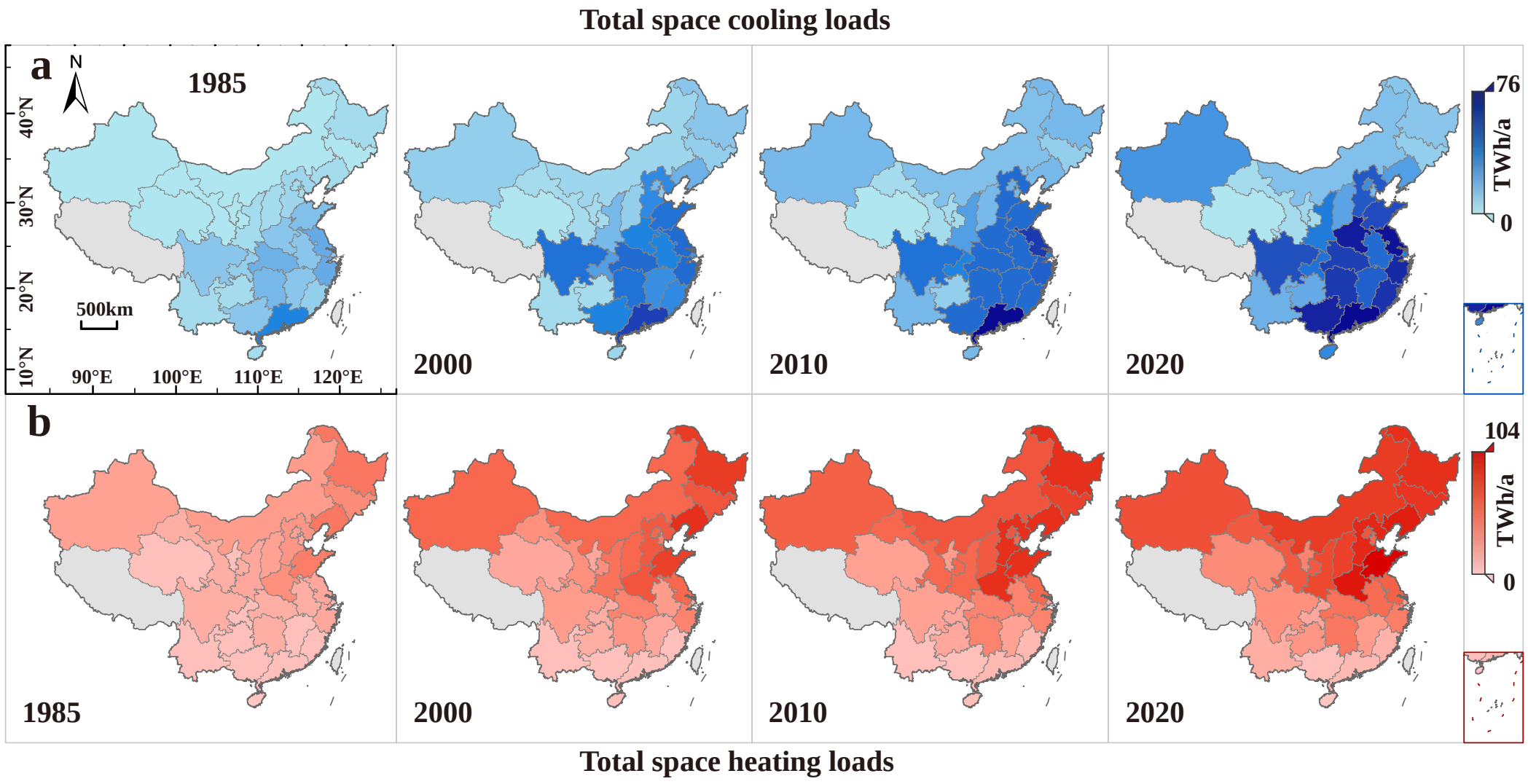

**Fig. 3.** Total space (a) cooling and (b) heating loads of urban residential buildings by province in 1985, 2000, 2010, and 2020.

Fig. 3b shows that the total heating load of urban residential buildings across Chinese provinces from 1985 to 2020 exhibited a distinct spatial pattern, with higher values concentrated in northwestern regions and lower values in southeastern regions, along with significant temporal variations. The spatial distribution reveals that heating load in 1985 was predominantly concentrated in northern zones, with only three provinces, namely, Liaoning (11.7 TWh/a), Heilongjiang (11.5 TWh/a), and Shandong (10.9 TWh/a), exceeding 10 TWh/a, whereas eleven southern provinces, including Hainan (0 TWh/a) and Yunnan (0.06 TWh/a), recorded less than 1 TWh/a, resulting in an 11.7 TWh/a difference between Liaoning, with the highest value, and Hainan, with the lowest value. This spatial heterogeneity intensified with increasing heating load. By 2000, thirteen provinces exceeded 10 TWh/a, with Liaoning (50.8 TWh/a) being the highest and Hainan (0 TWh/a) the lowest, yielding a 50.8 TWh/a differential. In 2010, fourteen provinces surpassed 10 TWh/a, while the gap between Liaoning (73.6 TWh/a) and Hainan ($0.48\times10^{-3}$ TWh/a) widened further to 73.6 TWh/a. In 2020, Shandong recorded the maximum heating load at 103.8 TWh/a, creating a 103.4 TWh/a differential with Hainan's $0.38\times10^{-3}$ TWh/a. Notably, all five provinces with the heaviest loads, namely, Shandong, Henan, Liaoning, Hebei, and Heilongjiang, were situated in zones of SC or C.

Temporal analysis reveals an accelerating growth trend in the national total heating load. The period from 1985 to 2000 marked a phase of rapid expansion, during which eleven provinces, including Liaoning (+39.1 TWh/a) and Heilongjiang (+35.8 TWh/a), demonstrated increases exceeding 10 TWh/a, whereas southern provinces, such as Hainan, maintained minimal growth. From 2000 to 2010, growth patterns became increasingly differentiated. Eight provinces sustained annual increases exceeding 10 TWh/a, particularly Shandong (+30.4 TWh/a) and Hebei (+25.2 TWh/a). In contrast, four provinces experienced negative growth, namely, Hainan, Hubei, Zhejiang and Sichuan. The period from 2010 to 2020 witnessed explosive growth in northern zones, with ten provinces, including Henan (+36.5 TWh/a) and Shandong (+31 TWh/a), achieving annual increases surpassing 10 TWh/a, whereas Hainan remained the sole province that experienced persistent negative growth. This spatiotemporal evolution substantiates the strong correlation between China's heating load and climatic zones while highlighting the intensifying thermal load pattern in northern zones with centralized heating systems.

As shown in Fig. 4a, the total cooling load of rural residential buildings across China's provinces exhibited a distinct spatial distribution characterized by higher values in the southeast and lower values in the northwest, with regional disparities progressively widening over time. In 1985, provincial cooling loads already demonstrated a clear gradient distribution, with 13 provinces exceeding 10 TWh/a, the majority of which were clustered within the 10-20 TWh/a range. Guangdong Province (34.0 TWh/a) ranked highest nationally as the only province surpassing 30 TWh/a, while Guangxi (28.8 TWh/a) and Hunan (22.4 TWh/a) exceeded 20 TWh/a. Five provinces recorded total cooling loads below 1 TWh/a, with Qinghai Province (0.26 TWh/a) being the lowest nationally, showing an extreme difference of 33.8 TWh/a compared with Guangdong. By 2000, the number of provinces exceeding 10 TWh/a remained at 13, where the 13th-ranked Jiangxi (17.5 TWh/a) substantially exceeded Chongqing (9.8 TWh/a), which fell below 10 TWh/a, whereas the maximum difference between the top-ranked Guangdong (45.9 TWh/a) and bottom-ranked Qinghai (0.41 TWh/a) reached 45.5 TWh/a. In 2010, 15 provinces surpassed 10 TWh/a in total cooling load, with Guangdong (50.9 TWh/a) showing a difference of 50.5 TWh/a from Qinghai (0.32 TWh/a), the lowest. By 2020, the provincial distribution reveals that over half of the provinces exceeded 10 TWh/a, including six provinces above 30 TWh/a, whereas seven provinces remained below 1 TWh/a. The extreme difference between highest-load Guangdong (63.0 TWh/a) and lowest-load Qinghai (0.14 TWh/a) increased to 62.9 TWh/a.

Longitudinal analysis reveals a growing divergence in rural cooling loads, with southern China exhibiting increasing loads and northern regions showing declining trends. During the period from 1985-2000, all provinces experienced growth in total cooling load, with southern provinces showing particularly significant increases—five provinces, including Guangxi and Henan—with recorded increases exceeding 10 TWh/a, whereas northern provinces, such as Qinghai, experienced only marginal growth (0.16 TWh/a). After 2000, regional divergence intensified: from 2000 to 2010, southern provinces maintained rapid growth, with Jiangxi, Hunan, and others registering increases of approximately 15 TWh/a, whereas 13 northern provinces exhibited declines, among which Henan showed the largest reduction (5.5 TWh/a). Between 2010 and 2020, southern provinces generally experienced slower growth, with only Guangdong sustaining an increase exceeding 10 TWh/a, whereas over half of the northern provinces continued to decline, with Hebei registering the most substantial decrease (6.5 TWh/a).

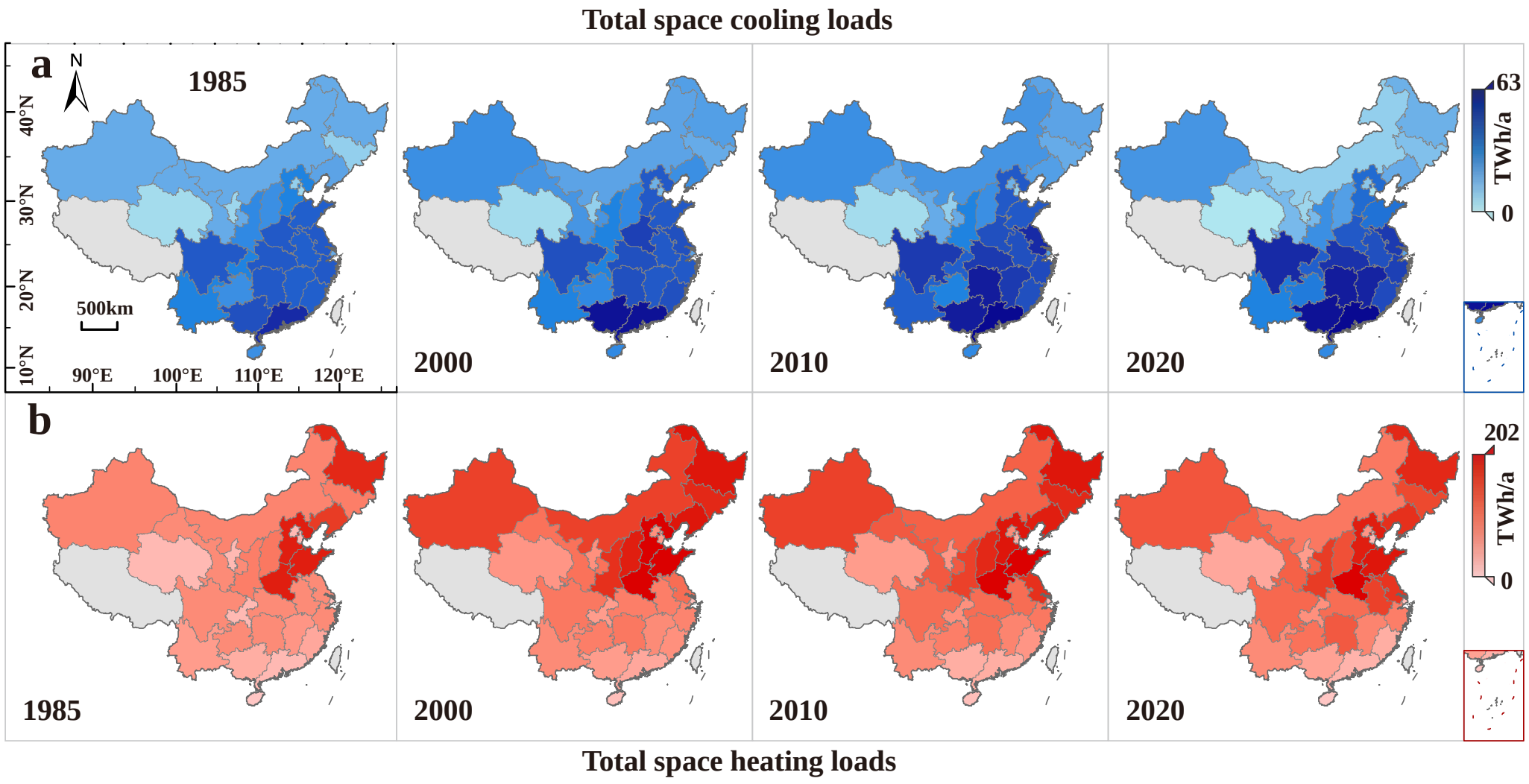

**Fig. 4.** Total space (a) cooling and (b) heating loads of rural residential buildings by province in 1985, 2000, 2010, and 2020.

As shown in Fig. 4b, the total heating load of rural residential buildings across China's provinces exhibited a distinct spatial distribution pattern characterized by higher values in the northwest and lower values in the southeast, with regional disparities initially expanding before gradually narrowing over time. In 1985, provincial heating loads already displayed significant variation, with Henan Province ranking highest (104.8 TWh/a) and Hainan having the lowest value ($3.1\times10^{3}$ TWh/a), resulting in an extreme difference of 104.8 TWh/a. By 2000, the gap between top-ranked Henan (201.6 TWh/a) and bottom-ranked Hainan (0.11 TWh/a) has widened to 201.5 TWh/a.

This disparity decreased to 181.6 TWh/a in 2010 between Henan (181.6 TWh/a) and Hainan (0.06 TWh/a) and further declined to 174.6 TWh/a by 2020 between Henan (174.6 TWh/a) and Hainan (0.05 TWh/a). Provincial distribution analysis reveals that in 1985, only Henan exceeded 100 TWh/a, five provinces surpassed 50 TWh/a, and ten provinces fell below 10 TWh/a. By 2000, five provinces (Henan, Shandong, Hebei, Heilongjiang, and Liaoning) exceeded 100 TWh/a, another five surpassed 50 TWh/a, and only three remained below 10 TWh/a. In 2010, while the number of provinces above 100 TWh/a remained unchanged, their values decreased, whereas those exceeding 50 TWh/ increased to seven, and those below 10 TWh/a rose again to ten. By 2020, only three provinces (Henan, Shandong, and Hebei) maintained heating loads above 100 TWh/a, fourteen exceeded 50 TWh/a, and seven southern provinces dropped below 10 TWh/a. Notably, across all four benchmark years, Hainan remained the sole province with a consistently sub-1 TWh/a heating load.

From a temporal perspective, China's rural heating load exhibited pronounced geographical redistribution, characterized by declining usage in the north and expanding load in the south. During the period from 1985 to 2000, all provinces exhibited growth in total heating load. The northern provinces showed more pronounced increases, where five provinces, including Henan and Shandong, recorded increases exceeding 50 TWh/a, whereas southern provinces, such as Hainan, experienced minimal growth of less than 0.1 TWh/a. After 2000, the regional pattern underwent significant transformation. From 2000 to 2010, the growth center shifted southward, with provinces such as Jiangsu and Anhui achieving increases exceeding 15 TWh/a, while all provinces in North China and Northeast China experienced declines, and Heilongjiang showed the largest reduction (20.9 TWh/a). During 2010-2020, the southwestern zone emerged as the primary growth area, with only Anhui maintaining an increase above 10 TWh/a, whereas 18 northern provinces and autonomous zones generally exhibited declines, with Liaoning registering the most dramatic decrease (33.1 TWh/a).

Fig. 5 shows the total cooling and heating loads of urban and rural residential buildings across 30 provinces in China from 1985-2024, revealing significant regional and urban-rural disparities as well as varying trends over time. Among the 30 provinces, distinct differences were observed between urban and rural cooling and heating loads. From 1985 to 2024, heating loads dominated in zones of SC and C, with urban heating load increasing substantially. In the HSCW zone, both the cooling and heating loads remained high, with urban cooling loads exhibiting particularly rapid growth. In contrast, both the zones of HSWW and M were characterized predominantly by cooling loads, with urban and rural heating loads significantly lower than the cooling loads and the urban heating load nearly negligible. Notably, rural cooling and heating loads surpassed urban levels in 27 provinces, with the exclusion of Beijing, Tianjin and Shanghai. In Beijing and Tianjin, urban cooling and heating loads were higher than those in rural areas, whereas Shanghai exhibited no significant urban-rural disparity in heating loads, although urban cooling loads remained higher than rural levels overall.

From 1985 to 2024, the total cooling and heating loads of both urban and rural residential buildings across all provinces exhibited fluctuating trends. For heating loads, the total rural residential heating load in most zones initially increased before it declined, whereas the urban residential heating loads demonstrated consistent growth throughout the study period. In the zones of SC and C, the total heating load of urban residential buildings was nearly negligible in 1985 and was significantly lower than that of rural buildings during the same period. Over time, urban heating loads exhibited consistent growth, whereas rural heating loads initially increased but then decreased. By 2024, urban and rural heating loads have largely converged in most provinces. Notably, Liaoning and Inner Mongolia experienced rapid growth in urban heating load, surpassing their respective rural levels as early as 2015 and 2017. In contrast, provinces such as Henan and Gansu presented slower declines in rural heating loads alongside modest urban growth. Consequently, urban heating loads have remained substantially lower than rural heating loads by 2024, albeit with a considerably narrowed gap since 1985. Beijing and Tianjin presented unique cases: in 1985, their rural heating loads (Beijing, 6.22 TWh/a; Tianjin, 5.35 TWh/a) were only marginally higher than their urban loads (Beijing, 3.75 TWh/a; Tianjin, 2.53 TWh/a), but their urban load (Beijing, 16.22 TWh/a; Tianjin, 10.89 TWh/a) exceeded their rural levels (Beijing, 12.78 TWh/a; Tianjin, 10.69 TWh/a) by 2000. In the zones of HSCW, HSWW, and M, rural heating loads persistently exceeded urban levels by a substantial margin throughout the study period. Shanghai, however, diverged from this pattern, exhibiting nearly identical urban and rural heating load magnitudes with parallel trends. This distinctive characteristic was not observed in other provinces.

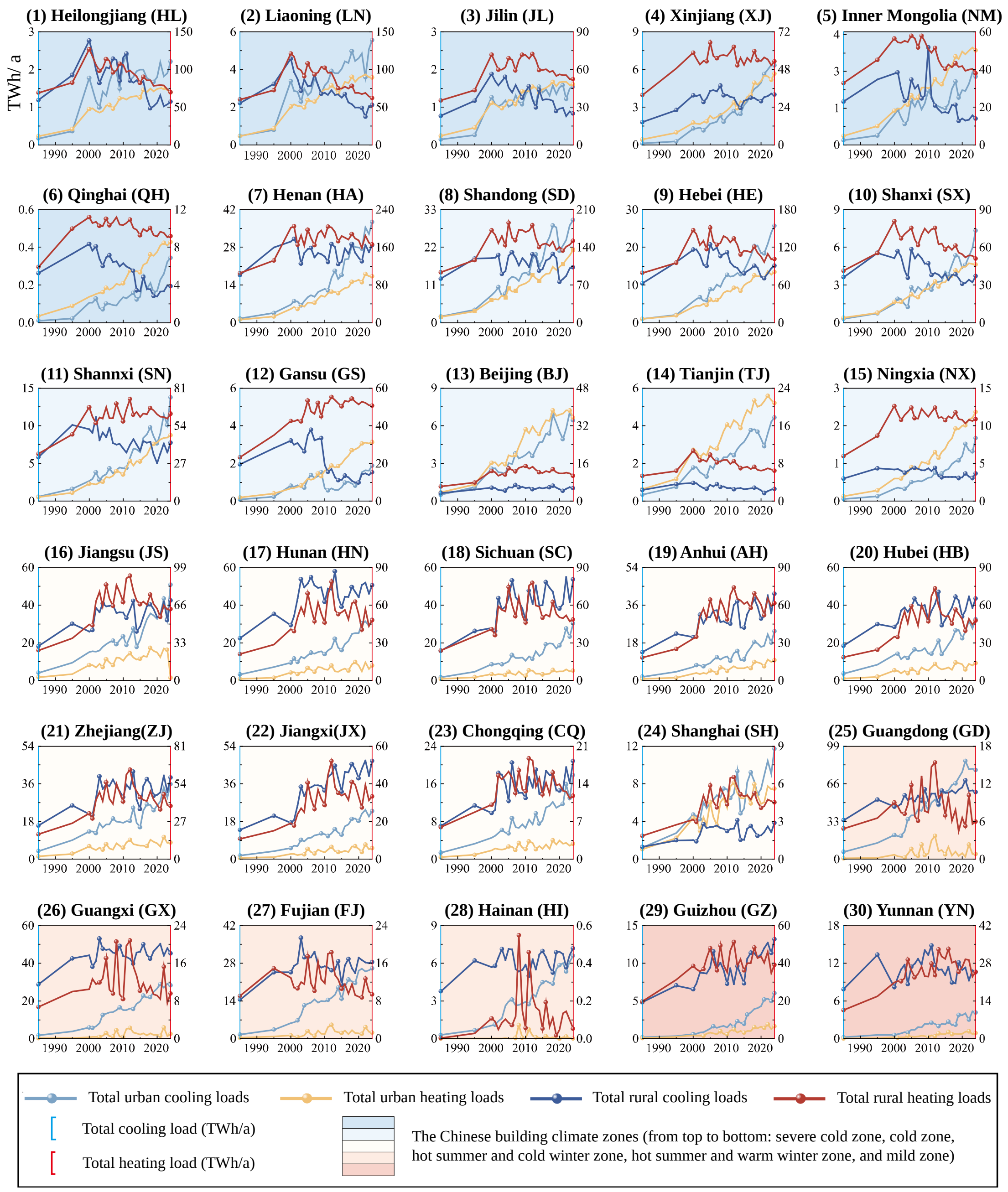

**Fig. 5.** Trends in space cooling and heating loads of urban and rural buildings by province, 1985-2024.

Significant urban-rural disparities were also observed in cooling loads. In the zones of SC and C, the rural cooling loads generally initially increased but then decreased, whereas the urban cooling loads steadily increased. The urban cooling loads surpassed the rural levels in most provinces approximately 2015. However, Gansu exhibited a notable lag, with urban cooling loads remaining below rural levels until 2020. It was not until 2021 that urban cooling loads (1.54 TWh/a) exceeded rural levels (1.42 TWh/a) in Gansu, after which the gap between them remained relatively small for the following three years. Beijing and Tianjin demonstrated distinct patterns, characterized by stable rural cooling load fluctuations and consistently higher urban cooling loads after 2000. In southern zones, rural cooling loads generally exceeded urban levels in HSCW zones, with Shanghai being the exception. In Shanghai, urban cooling loads consistently remained higher than rural cooling loads did, with an expanding disparity over time. Compared with rural areas, the HSWW zones presented faster growth in urban cooling loads, leading to a gradual reduction in the urban-rural gap. Guangdong represented a notable case where urban cooling loads (59.12 TWh/a) overtook rural levels (58.09 TWh/a) in 2015. In the M zone, both urban and rural cooling loads showed increasing trends, with rural levels consistently maintaining higher values than urban loads throughout the study period. Overall, China's urban and rural residential building cooling and heating loads have undergone profound transformations over the past four decades. In most provinces, the rapid growth of urban loads relative to the slower increase in rural loads resulted in a general narrowing of the gap between the total heating and cooling loads of urban and rural residential buildings. This trend also showed distinct evolutionary trajectories across different climate zones in the north and south.

Section 4.1 presents the total cooling and heating loads and their temporal trends in China's urban and rural residential buildings, revealing the urban-rural disparities in thermal load and regional variations in load requirements, which partially addresses the first research question raised in Section 1.

### *4.2. Changes in residential cooling and heating loads per floorspace with urban-rural disparities*

Figs. 6 and 7 present the spatiotemporal characteristics of cooling and heating loads per floorspace in urban and rural residential buildings across 30 Chinese provinces from 1980-2024. The results demonstrate pronounced spatial variations and temporal trends in regional cooling and heating loads. Analysis reveals that cooling loads in both urban and rural buildings were predominantly concentrated in the zones of HSCW and HSWW, whereas heating loads were primarily distributed across the zones of SC and C, with significantly higher values than those observed in the HSCW, HSWW, and M zones. This spatial pattern was fundamentally attributed to the climatic conditions of each province.

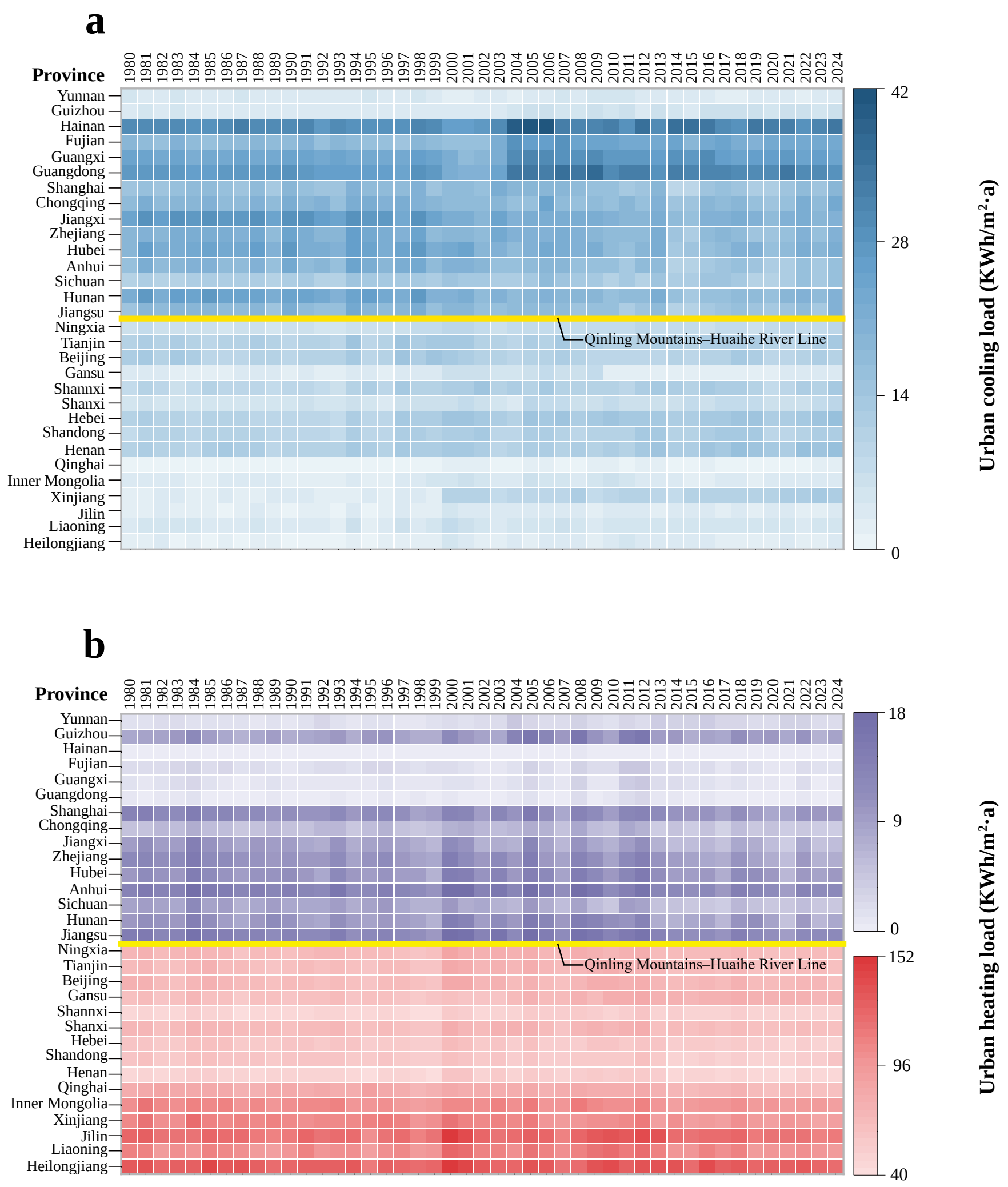

**Fig. 6.** Per floorspace (a) cooling load and (b) heating load in urban China, 1980-2024.

From 1980 to 2024, China's urban areas exhibited a significant increasing trend in cooling load per floorspace, with the overall level rising from 12.4 kilowatt-hours per square meter per year (kWh/m$^2$/a) to 15.1 kWh/m$^2$/a, representing an average annual growth rate of 0.44%. As illustrated in Fig. 6a, distinct gradient distribution patterns were observed across climate zones. The HSWW zone consistently recorded the highest cooling loads, with provinces such as Guangdong (increasing from 27.73 kWh/m$^2$/a in 1980 to 29.71 kWh/m$^2$/a in 2024), Fujian (24.57 kWh/m$^2$/a to 25.40 kWh/m$^2$/a), and Hainan (31.09 kWh/m$^2$/a to 33.94 kWh/m$^2$/a) being particularly prominent. The HSCW zone, including Jiangxi (24.99 kWh/m$^2$/a to 21.34 kWh/m$^2$/a) and Hubei (19.34 kWh/m$^2$/a to 22.19 kWh/m$^2$/a), formed the second tier, whereas the C zone such as Beijing (12.28 kWh/m$^2$/a to 11.79 kWh/m$^2$/a) and Hebei (11.55 kWh/m$^2$/a to 16.38 kWh/m$^2$/a) ranked third. In contrast, SC and M zones maintained the lowest cooling loads nationwide. Specifically, in SC zone, Heilongjiang had values ranging from 2.72 to 3.80 kWh/m$^2$/a, Inner Mongolia had values ranging from 3.94 to 4.36 kWh/m$^2$/a, and Qinghai had values varying between 1.12 and 2.66 kWh/m$^2$/a. Similarly, in the M zone, Guizhou reported values ranging from 3.74 to 7.36 kWh/m$^2$/a, whereas Yunnan reported values ranging from 4.94 to 4.52 kWh/m$^2$/a. Temporally, most zones displayed fluctuating upward trends after 2000, with notable increases observed in some areas in the SC and C zones, such as Xinjiang (from 3.25 kWh/m$^2$/a in 1980 to 12.79 kWh/m$^2$/a in 2024) and Henan (10.53 kWh/m$^2$/a to 16.98 kWh/m$^2$/a). Despite limited overall changes in cooling loads over the 40-year period, interannual variability was evident across most zones.

During the period from 1980-2024, China's urban areas demonstrated a significant decreasing trend in heating load per floorspace, with the national average decreasing from 44.08 kWh/m$^2$/a to 39.92 kWh/m$^2$/a at an annual rate of 0.23%. Fig. 6b shows distinct spatial variations in heating loads across different zones, which are attributed primarily to climatic conditions and heating load disparities. Compared with other areas, the SC and C zones presented substantially higher heating loads because of their extended heating seasons and greater load. In the SC zone, Heilongjiang (decreasing from 132.20 kWh/m$^2$/a in 1980 to 119.59 kWh/m$^2$/a in 2024) and Jilin (122.47 kWh/m$^2$/a to 111.85 kWh/m$^2$/a) recorded the highest values, with averages exceeding 120 kWh/m$^2$/a, followed by Liaoning (107.25 kWh/m$^2$/a to 95.33 kWh/m$^2$/a), Xinjiang and Inner Mongolia, all maintaining averages above 100 kWh/m$^2$/a. The C zones such as Shanxi (73.47 kWh/m$^2$/a to 63.31 kWh/m$^2$/a), Beijing (74.21 kWh/m$^2$/a to 65.05 kWh/m$^2$/a) and Henan (47.86 kWh/m$^2$/a to 44.76 kWh/m$^2$/a) presented moderate heating loads of approximately 50 kWh/m$^2$/a. In contrast, the HSCW zones such as Jiangsu (13.67 kWh/m$^2$/a to 11.46 kWh/m$^2$/a) and Sichuan (8.09 kWh/m$^2$/a to 4.43 kWh/m$^2$/a) displayed much lower values of approximately 8 kWh/m$^2$/a, whereas HSWW zones such as Guangdong (8.09 kWh/m$^2$/a to 4.43 kWh/m$^2$/a) registered minimal loads of approximately 1 kWh/m$^2$/a. The M zone exhibited notable internal variations, with Yunnan (1.17 kWh/m$^2$/a to 2.16 kWh/m$^2$/a) showing the nation's lowest value at 2 kWh/m$^2$/a, whereas Guizhou (7.94 kWh/m$^2$/a to 7.96 kWh/m$^2$/a) presented a relatively high value of 9 kWh/m$^2$/a. Temporally, the heating loads exhibited a general declining trend nationwide, although with marked regional differences. The most significant reductions occurred in the SC (Heilongjiang, Jilin) and C zones (Liaoning, Inner Mongolia), with decreases reaching 11.24% and 5.18%, respectively. In contrast, the HSCW, HSWW and M zones experienced relatively smaller cumulative declines (21.4%) because of their initially low baseline values. These patterns effectively reflected the evolving characteristics of building heating loads across different climate zones.

From 1980 to 2024, the floorspace cooling load of rural residential buildings in China showed a declining trend, with the national average decreasing from 22.63 kWh/m$^2$/a to 19.87 kWh/m$^2$/a, at an average annual rate of -0.29%. Fig. 7a illustrates the distinct climate zone-specific evolution of the cooling load per floorspace in rural China from 1980-2024. Spatially, the cooling loads demonstrated a clear hierarchical pattern across climate zones: the HSWW zone consistently presented the highest values nationwide, with Guangdong (77.11 kWh/m$^2$/a in 1980 to 39.75 kWh/m$^2$/a in 2024), Hainan (82.36 kWh/m$^2$/a to 47.14 kWh/m$^2$/a), and Guangxi (62.45 kWh/m$^2$/a to 34.85 kWh/m$^2$/a) being particularly prominent. The HSCW zone, represented by Jiangxi (28.25 kWh/m$^2$/a to 36.09 kWh/m$^2$/a) and Hubei (23.02 kWh/m$^2$/a to 37.21 kWh/m$^2$/a), formed the secondary tier, whereas the M zone, including Yunnan (22.79 kWh/m$^2$/a to 10.08 kWh/m$^2$/a) and Guizhou (12.12 kWh/m$^2$/a to 13.32 kWh/m$^2$/a), along with the C zone exemplified by Beijing (16.83 kWh/m$^2$/a to 9.91 kWh/m$^2$/a) and Hebei (16.75 kWh/m$^2$/a to 12.88 kWh/m$^2$/a), ranked third. The SC zone, encompassing Xinjiang (14.19 kWh/m$^2$/a to 11.10 kWh/m$^2$/a), Inner Mongolia (11.32 kWh/m$^2$/a to 3.75 kWh/m$^2$/a), and Liaoning (8.47 kWh/m$^2$/a to 5.28 kWh/m$^2$/a), persistently maintained the lowest cooling loads nationwide. Temporally, divergent trends emerged among climate zones: the HSCW zone showed a sustained increase in the cooling load from 21.13 kWh/m$^2$/a in 1980 to 34.04 kWh/m$^2$/a in 2024, indicating substantially increased cooling load, whereas the HSWW zone displayed the most pronounced decline, with mean values decreasing from 66.23 to 38.27 kWh/m$^2$/a, reaching levels comparable to those of the HSCW zone by 2020. Other zones generally exhibited fluctuating downward trends, with Yunnan in the M zone and parts of the C zone demonstrating particularly notable reductions after 2000. Notably, although absolute changes in cooling loads over the four decades remained relatively modest in most zones, all areas showed marked interannual variability, suggesting the combined effects of climatic variations and anthropogenic adaptation measures.

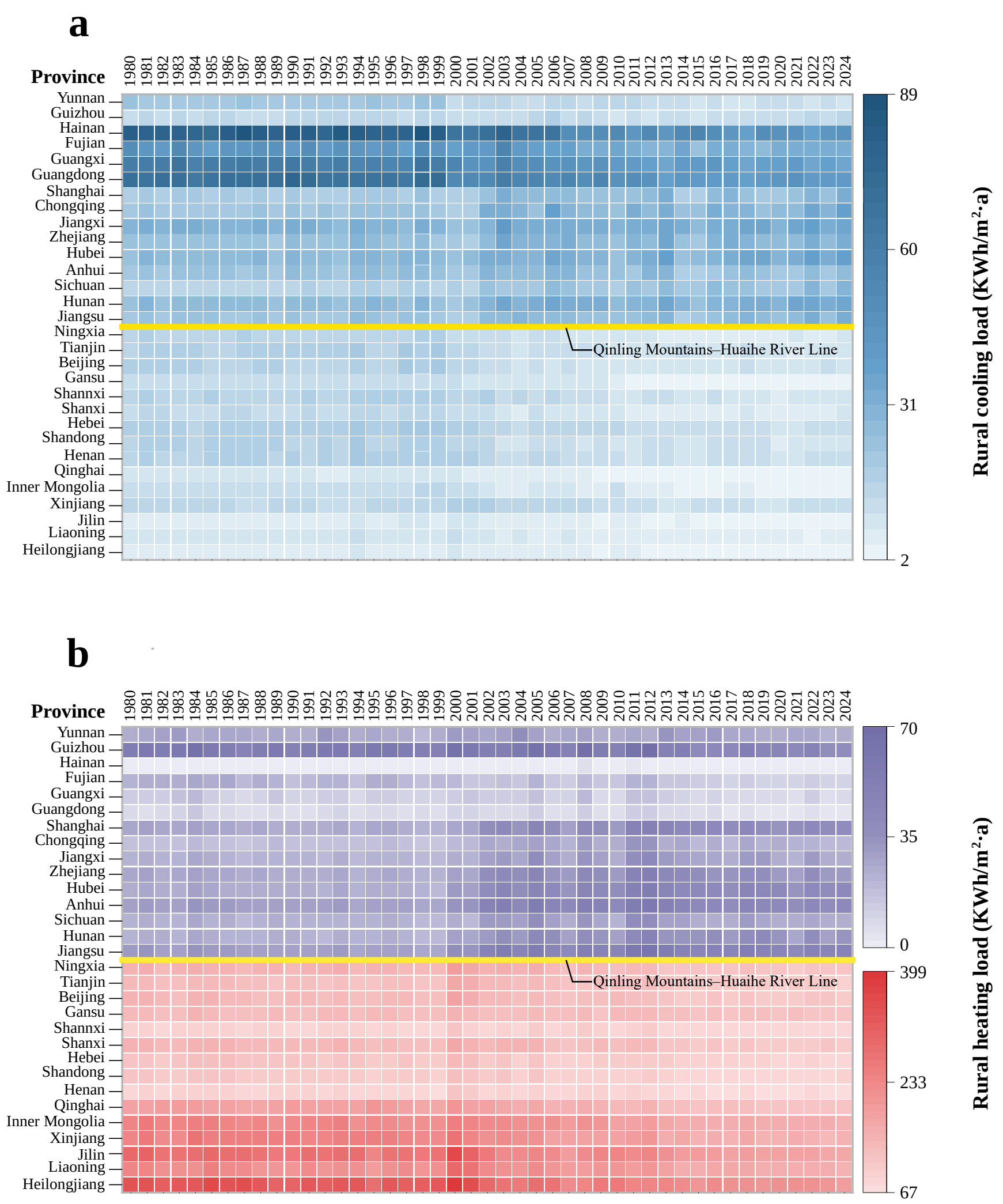

**Fig. 7.** Per floorspace (a) cooling load and (b) heating load in rural China, 1980-2024.

From 1980 to 2024, the floorspace heating load of rural residential buildings in China exhibited a downward trend, with the national average decreasing from 100.15 kWh/m$^2$/a to 72.42 kWh/m$^2$/a, at an average annual rate of -0.73%. Fig. 7b shows the pronounced regional heterogeneity in the heating load per floorspace of rural residential buildings in China from 1980-2024. The analysis reveals a general declining trend across the zones of SC, C, HSWW, and M, whereas the HSCW exhibited a distinctive increasing pattern, reflecting the differential evolution of heating loads among climate zones. Spatially, heating loads displayed a clear hierarchical distribution: zones with SC and C presented significantly higher values due to extended heating periods and greater load. Heilongjiang (339.91 kWh/m$^2$/a in 1980 to 197.58 kWh/m$^2$/a in 2024) emerged as the province with the highest rural heating load nationwide, forming the primary tier alongside other SC zones, including Jilin (291.92 kWh/m$^2$/a to 176.11 kWh/m$^2$/a), Liaoning (237.19 kWh/m$^2$/a to 155.28

kWh/m$^2$/a), and Inner Mongolia (241.02 kWh/m$^2$/a to 152.65 kWh/m$^2$/a), all maintaining average values exceeding 150 kWh/m$^2$/a. The C zone such as Shanxi (153.01 kWh/m$^2$/a to 107.53 kWh/m$^2$/a), Beijing (148.42 kWh/m$^2$/a to 103.08 kWh/m$^2$/a), and Ningxia (151.36 kWh/m$^2$/a to 114.59 kWh/m$^2$/a) presented intermediate levels of approximately 100 kW/m$^2$/a. Comparatively lower heating loads characterize two regions: In the M zone, Guizhou decreased from 54.45 to 39.57 kWh/m$^2$/a, and Yunnan declined slightly from 25.28 to 23.60 kWh/m$^2$/a, whereas in the HSCW zone, Anhui increased from 29.91 to 41.07 kWh/m$^2$/a, and Zhejiang rose from 27.21 to 32.53 kWh/m$^2$/a. The HSWW zone demonstrated minimal values, particularly in Guangxi (14.08 kWh/m$^2$/a to 7.40 kWh/m$^2$/a) and Fujian (23.27 kWh/m$^2$/a to 10.34 kWh/m$^2$/a), with Hainan (0.47 kWh/m$^2$/a to 0.34 kWh/m$^2$/a) approaching zero, the lowest national level. Temporally, the national rural heating load showed an overall decreasing trend, with the most substantial reductions occurring after 2000 in the SC (Heilongjiang, Jilin) and C zones (Liaoning, Inner Mongolia), followed by the M zone. Despite their initially low baselines, the HSWW zone maintained a gradual decline. Notably, the HSCW zone displayed a countertrend increase, potentially attributable to growing heating load and elevated thermal comfort expectations under climate change. These spatially differentiated evolutionary patterns not only elucidate the profound influence of climatic conditions on building thermal load but also provide critical evidence for formulating zone-specific energy conservation policies for rural buildings. Section 4.2 analyzes the per floorspace cooling and heating load dynamics in China's urban and rural residential buildings. Taken together, the findings in Sections 4.1 and 4.2 provide a complete solution to the first research question posed in Section 1.

## 5. Discussion

### *5.1. Key drivers affecting urban-rural disparities in cooling and heating loads*

Eqs. (1-2) demonstrate that regional heating and cooling loads can be quantified as the product of the building stock and its corresponding energy efficiency level across different periods. The variation in loads essentially stems from the combined effects of the spatial distribution of building stocks ($A_x$) constructed during different eras and their prototype-specific load intensities per floorspace ($q_{Hx}$). Fig. 8 further illustrates the different drivers of the urban-rural disparities in heating and cooling loads. Specifically, the scale of the building stock and its spatiotemporal patterns directly determine the total thermal load, whereas the thermal performance of building envelopes in prototypes governs the load intensity of individual structures. These two categories of factors are influenced by urbanization processes and building energy efficiency standards: Urbanization reshapes building stock structures by altering land-use patterns, whereas energy efficiency standards regulate envelope design criteria to define the energy performance characteristics of building prototypes. This divergence in built environments, driven jointly by macrolevel socioeconomic factors and technical regulatory frameworks, ultimately leads to differentiated space heating and cooling loads between urban and rural areas.

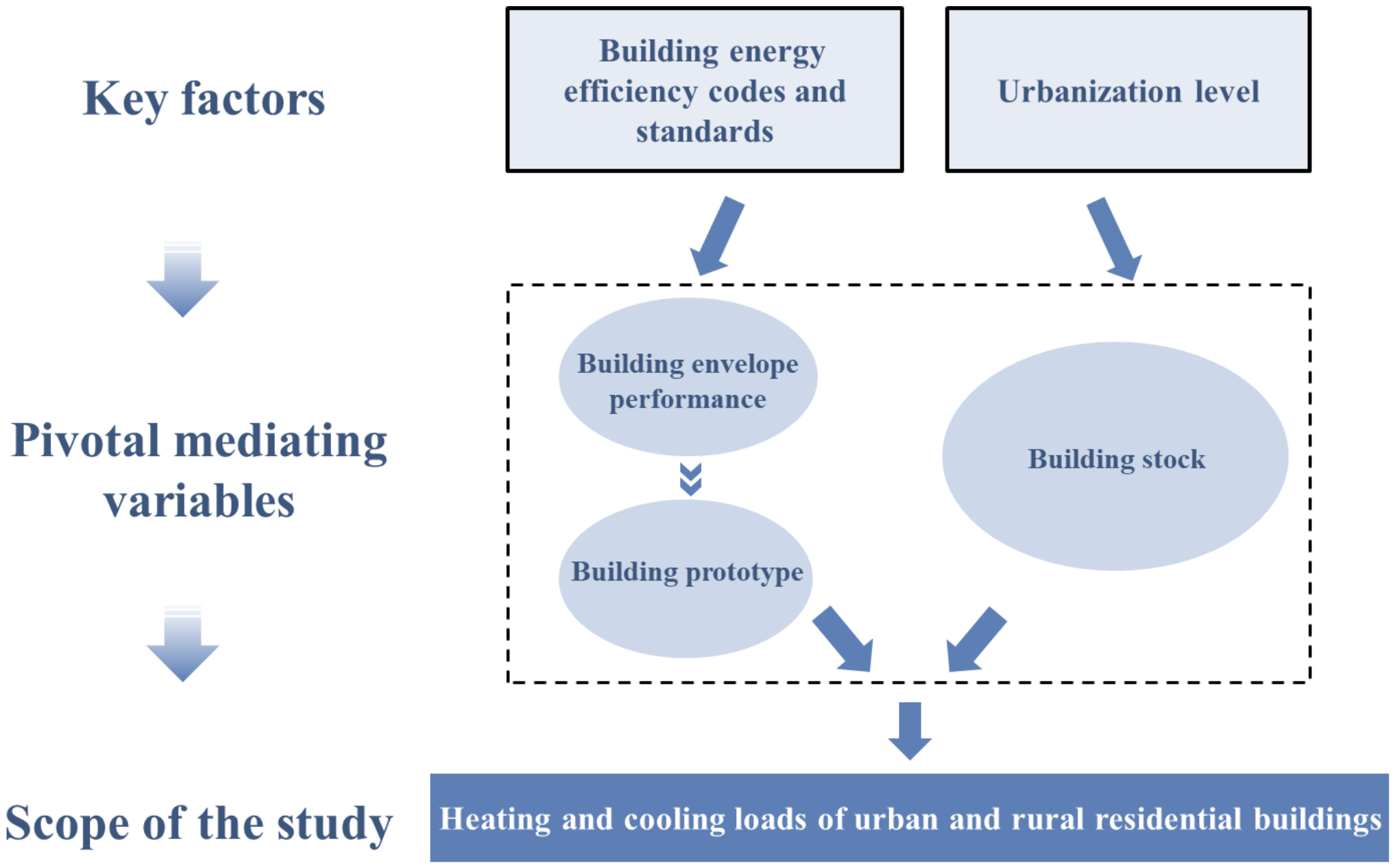

**Fig. 8.** Framework illustrating the key factors in cooling and heating loads for urban and rural areas.

To elucidate the underlying mechanisms of urban-rural disparities in heating and cooling loads, this study conducts a quantitative analysis through a series of figures. Fig. 9 illustrates the evolution of residential building stocks and their corresponding load variations during urbanization. Fig. 10 selects representative provinces from each building thermal climate zone to compare urban and rural residential building prototypes across different periods. Furthermore, Fig. 11 examines the differences in envelope performance between urban and rural prototypes from an institutional perspective by comparing building energy efficiency standards. These findings collectively validate the dual-pathway mechanism proposed in our framework: (1) urbanization-induced building stock expansion and (2) the energy efficiency standards governing building prototypes jointly explain urban-rural thermal load disparities. This provides empirical evidence for the observed divergence in space heating and cooling loads.

Fig. 9 illustrates the temporal evolution of the building stock across 30 Chinese provinces, along with the variations in residential cooling and heating loads in urban and rural areas during typical years. Specifically, Fig. 9a delineates the temporal trends of the total residential building stock in both urban and rural areas across these provinces from 2000-2024. Fig. 9b presents a comparative analysis of residential building stock constructed during different periods (as classified in Table 1 by construction year) and the corresponding total cooling and heating loads for three representative years (2005, 2015, and 2024). The color scheme employs purple and blue to distinguish urban and rural residential building stocks, respectively. Furthermore, the background gradient from blue to red indicates the provincial climatic zones, sequentially representing the SC, C, HSCW, HSWW, and M zones, in accordance with China's building thermal zoning standards. This visualization effectively captures the spatial-temporal characteristics of the building stock distribution and its associated cooling and heating loads across different climatic zones.

The data analysis presented in Fig. 9a reveals significant disparities in the urban-rural building stock distributions across Chinese provinces in 2000. During this baseline year, the rural residential building stock exceeded the urban stock in most provinces. Only four provincial-level administrative units exhibited the opposite pattern: Beijing, Tianjin, Shanghai, and Guangdong. Notably, in Guangdong, the urban building stock reached 930 million $m^2$, slightly surpassing the rural stock of 867 million $m^2$ by a narrow margin of 63 million $m^2$. By 2024, this urban-rural distribution pattern has undergone substantial transformation. The urban residential building stock exceeded the rural stock in 22 provinces nationwide, whereas the remaining eight provinces (Henan, Gansu, Sichuan, Anhui, Jiangxi, Guangxi, Guizhou, and Yunnan) maintained but reduced rural predominance. The specific distributions were as follows: Henan (urban 2202 million $m^2$/rural 2284 million $m^2$), Gansu (urban 423 million $m^2$/rural 427 million $m^2$), Sichuan (urban 1761 million $m^2$/rural 1911 million $m^2$), Anhui (urban 1360 million $m^2$/rural 1499 million $m^2$), Jiangxi (urban 1078 million $m^2$/rural 1303 million $m^2$), Guangxi (urban 1113 million $m^2$/rural 1297 million $m^2$), Guizhou (urban 817 million $m^2$/rural 991 million $m^2$), and Yunnan (urban 921 million $m^2$/rural 1051 million $m^2$).

Three distinct evolutionary patterns emerged during the study period (2000-2024): 1) Persistent urban-dominant type: Exemplified by Beijing, Tianjin, and Shanghai, these zones maintained continuous urban superiority with nearly stagnant rural growth, resulting in widening urban-rural disparities. 2) Transitional type: Most provinces underwent urban-to-rural stock inversion, followed by accelerating divergence post-transition. 3) Persistent rural-dominant type: The eight aforementioned provinces exhibited two sub-patterns: an inverted U-curve of initial divergence followed by convergence or sustained narrowing trends. In conclusion, compared with 2000, all the provinces demonstrated significantly reduced urban-rural disparities. This spatial evolution reflects the dual dynamics of China's urbanization: the urban building stock expands continuously through population agglomeration, whereas rural stock contracts due to outmigration and resource reallocation, marking a systemic restructuring of China's urban-rural spatial organization. Crucially, even in provinces retaining rural advantages, the diminishing predominance signals a new developmental phase in urban-rural relations.

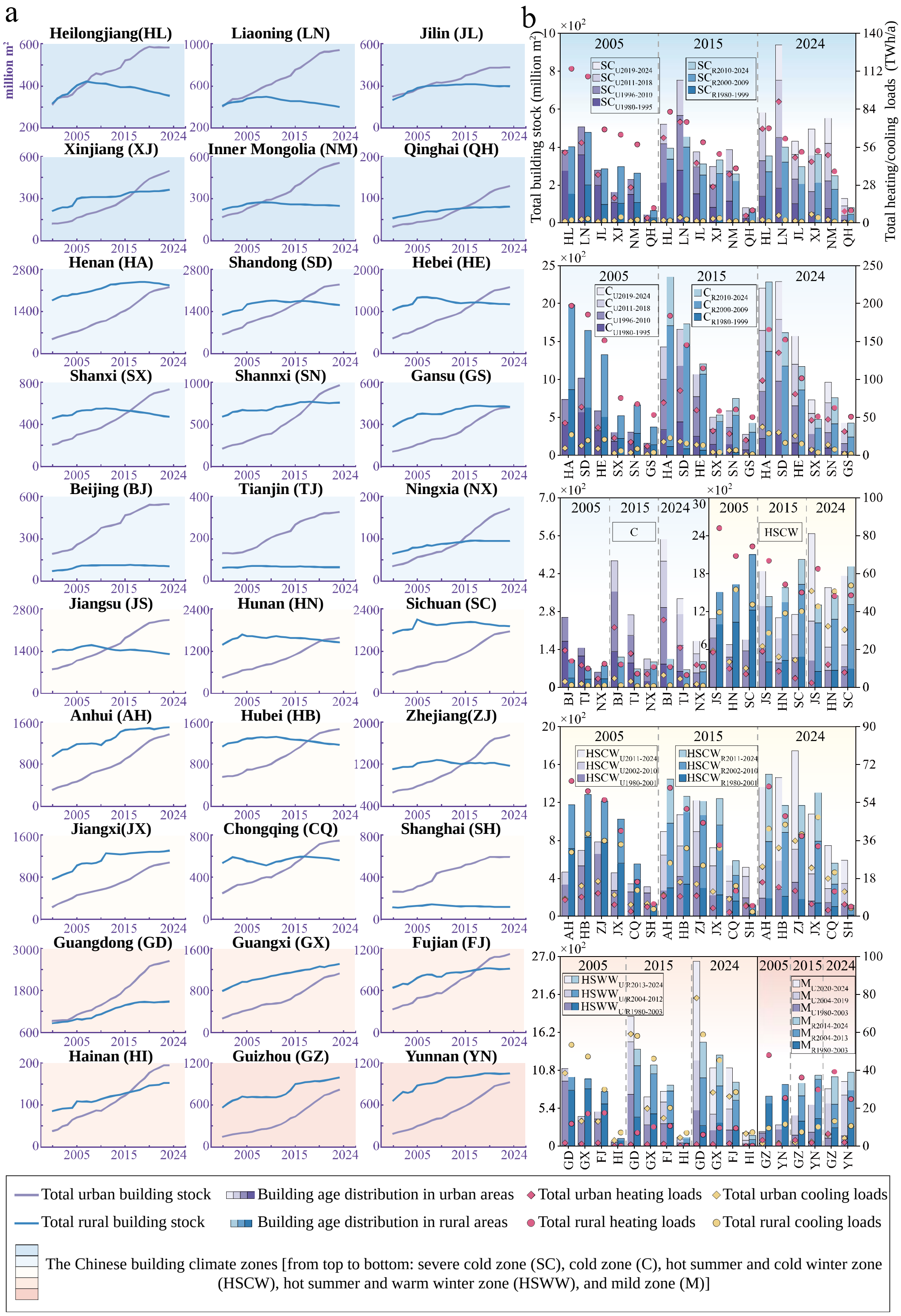

**Fig. 9.** Trends in residential building stock and heating/cooling loads in urban and rural China by province: (a) changes in urban and rural building stock, 2000-2024; (b) changes in building stock and corresponding heating/cooling loads in 2005, 2015, and 2024.

The analysis of Fig. 9b reveals distinct evolutionary trends in urban and rural residential cooling and heating loads across China from 2005-2024. During this period, the stock of newly constructed urban and rural residential buildings increased after 2010, whereas the stock of buildings constructed before 2010 (particularly those built prior to 2000) gradually declined. On the basis of disparities in building stock, the cooling and heating load dynamics were categorized into three patterns. In persistent urban-dominant zones such as Beijing, Tianjin, and Shanghai, urban cooling and heating loads consistently exceeded rural values, with widening gaps over time. For example, Beijing's urban cooling and heating loads in 2005 were 3.0 TWh/a and 19.5 TWh/a, respectively, whereas the rural loads were 1.1 TWh/a and 13.9 TWh/a, yielding urban-rural differences of 1.9 TWh/a (cooling) and 5.6 TWh/a (heating). By 2015, urban loads rose to 4.7 TWh/a (cooling) and 31.6 TWh/a (heating), whereas rural loads declined to 1.0 TWh/a and 12.0 TWh/a, expanding the differences to 3.7 TWh/a and 19.6 TWh/a, respectively. This trend continued in 2024, with urban loads further increasing to 6.4 TWh/a (cooling) and 35.6 TWh/a (heating), whereas rural loads dropped to 1.0 TWh/a and 10.7 TWh/a, resulting in even larger disparities (5.4 TWh/a cooling, 24.9 TWh/a heating). Similar trends were observed in Tianjin and Shanghai, where urban-rural load differences increased significantly over time. The transitional zones, exemplified by Shanxi Province, exhibited a shift in cooling load dominance. In 2005, Shanxi's urban cooling and heating loads were 2.8 TWh/a and 22.6 TWh/a, respectively, against rural loads of 5.9 TWh/a and 75.4 TWh/a, with initial differences of -3.1 TWh/a (cooling) and -52.8 TWh/a (heating). By 2015, urban cooling loads have increased to 4.1 TWh/a and heating loads to 32.1 TWh/a, whereas rural loads have decreased to 3.9 TWh/a and 58.3 TWh/a, narrowing the differences to 0.2 TWh/a (cooling) and -26.2 TWh/a (heating), respectively. By 2024, urban cooling loads (7.3 TWh/a) surpassed rural levels (3.7 TWh/a), reversing the earlier trend, with a difference of 3.6 TWh/a, whereas the heating load gap (-4.7 TWh/a) continues to shrink despite remaining rural dominated. In persistent rural-dominant zones such as Yunnan, rural cooling and heating loads remained higher than urban values did, although the differences gradually diminished. In 2005, Yunnan's urban cooling and heating loads were 1.4 TWh/a and 0.8 TWh/a, respectively, whereas the rural loads were 11.5 TWh/a and 25.3 TWh/a, yielding differences of -10.1 TWh/a (cooling) and -24.5 TWh/a (heating), respectively. By 2015, urban loads have increased to 2.3 TWh/a (cooling) and 1.7 TWh/a (heating), whereas rural loads have increased to 10.1 TWh/a and 29.7 TWh/a, reducing the gaps to -7.8 TWh/a and -28.0

TWh/a, respectively. By 2024, urban loads reached 4.2 TWh/a (cooling) and 2.0 TWh/a (heating), yet rural loads (10.6 TWh/a cooling, 24.8 TWh/a heating) retained their advantage, albeit with further diminished differences (-6.4 TWh/a cooling, -22.8 TWh/a heating).

This study demonstrates that urban-rural building stock disparities are key determinants of cooling and heating load distributions. In urban-dominant provinces, load differences expanded continuously; in transitional zones, cooling load gaps followed an inverted U curve (first decreasing but then increasing), whereas heating load differences steadily converged; and in rural-dominant provinces, load gaps contracted alongside shrinking stock disparities. Overall, a significant positive correlation exists between urban-rural building stock differences and cooling and heating load disparities from 2005-2024, reflecting the profound impact of urbanization on cooling and heating load patterns.

The thermal performance of building envelopes in China strictly adheres to the climatic zoning standards for building energy efficiency. This study selected representative provinces from five climatic zones as research subjects: Heilongjiang (SC), Beijing (C), Chongqing (HSCW), Guangdong (HSWW), and Yunnan (M). As illustrated in Fig. 10, analysis of the cooling and heating loads per floorspace in urban and rural buildings constructed during different periods in 2024 reveals a significant downward trend in thermal loads across all climatic zones as construction years advanced. In the SC and C zones, heating load dominated the thermal load patterns of buildings in Heilongjiang and Beijing, where the urban-rural disparity was particularly pronounced. According to the correspondence between construction periods and energy efficiency standards (urban buildings from 1980-1995 followed the U1986 standard, equivalent to rural buildings from 1980-1999 under R1995; urban 1996-2010 buildings implemented U1995, corresponding to rural 2000-2009 under R1999; urban 2011-2018 buildings adopted U2010; and 2019-2024 buildings used U2018, matching rural 2010-2024 buildings under R2009), the data reveal significant differences in thermal loads between urban and rural areas. Specifically, U1986 standard buildings in Heilongjiang presented a heating load of 132.0 kWh/m$^2$/a in 2024, whereas rural R1995 standard buildings reached 344.7 kWh/m$^2$/a, representing more than double the urban value. Although the gap narrowed in the latest standard period (U2018 at 111.7 kWh/m$^2$/a versus R2009 at 137.2 kWh/m$^2$/a), rural heating loads remained higher. Beijing displayed a similar pattern, with U1986-standard urban buildings showing 73.6 kWh/m$^2$/a compared with rural R1995's 153.4 kWh/m$^2$/a.

By the U2018 (59.8 kWh/m$^2$/a) and R2009 (91.6 kWh/m$^2$/a) phases, the urban-rural difference has decreased, whereas rural heating loads have maintained elevated levels.

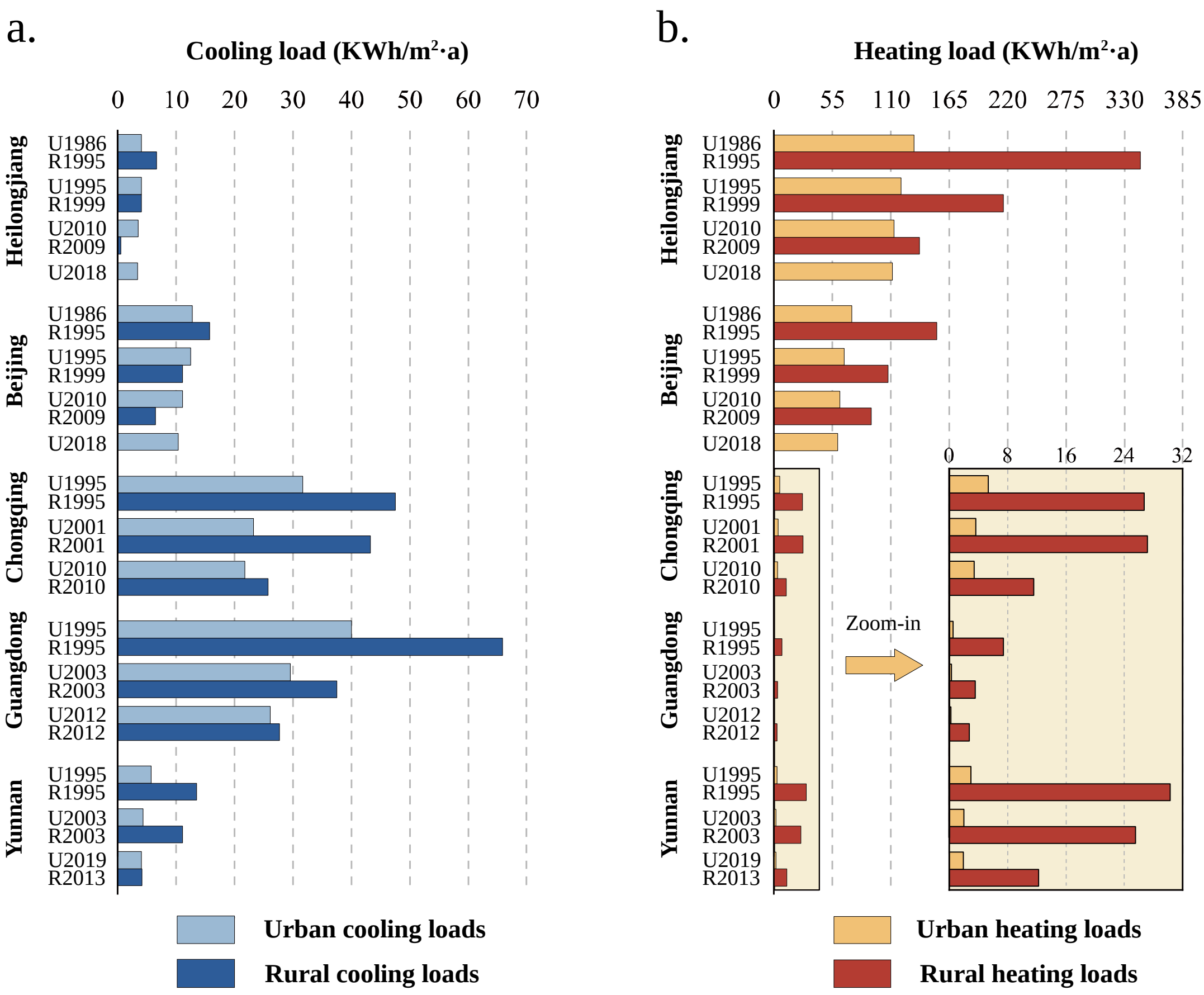

Note: U2019 represents the building energy efficiency standards for urban residential buildings issued in 2019; R2013 represents a baseline year of 2013 defined in this study for rural residential buildings.

**Fig. 10.** Decomposition of (a) cooling loads and (b) heating loads of urban and rural residential buildings by construction period (observed in 2024).

Chongqing in the HSCW zone shows more complex characteristics. U1995-standard urban buildings presented cooling and heating loads of 31.6 kWh/m$^2$/a and 5.3 kWh/m$^2$/a, respectively, whereas contemporaneous rural R1995 buildings reached 47.0 kWh/m$^2$/a and 26.7 kWh/m$^2$/a, with particularly notable differences in heating loads. Remarkably, R2001-standard rural buildings showed an anomalous increase in the heating load to 27.2 kWh/m$^2$/a, creating a wider gap than did urban U2001 buildings (3.6 kWh/m$^2$/a). Although subsequent standards (U2010/R2010) reduced overall thermal loads, rural values consistently exceeded urban values, highlighting the urgent need for improved winter insulation in rural buildings in this zone. In Guangdong, where cooling loads predominated, U1995 standard urban buildings recorded 40.1 kWh/m$^2$/a, whereas rural R1995 buildings recorded 65.9 kWh/$m^2$/a. As standards progressed through U2003/R2003 (29.5/37.5 kWh/$m^2$/a) and U2012/R2012 (26.1/27.7 kWh/$m^2$/a), the urban-rural disparity gradually diminished, although rural cooling loads remained persistently high. Monitoring data from Yunnan in the M zone indicated that while rural heating loads substantially decreased from 30.3 kWh/$m^2$/a under R1995 (compared with urban U1995's 3.0 kWh/$m^2$/a) to 25.5 kWh/$m^2$/a under R2003 (versus urban U2003's 2.0 kWh/$m^2$/a) and 12.2 kWh/$m^2$/a under R2013 (versus urban U2019's 1.9 kWh/$m^2$/a), they still significantly exceeded urban values.

A comprehensive analysis across climatic zones reveals that China's rural buildings consistently demonstrate higher heating loads per floorspace than urban structures do, particularly in zones of SC, C, HSCW, and M. Although evolving energy efficiency standards have gradually narrowed the urban-rural gap, the persistent underperformance of rural buildings in terms of thermal regulation necessitates targeted energy retrofit strategies. While the HSWW zone primarily exhibits cooling loads with relatively smaller urban-rural differences, rural energy efficiency improvements remain imperative. These findings provide crucial empirical evidence for implementing differentiated urban-rural building energy conservation strategies in China. Further decoupling and factor-contribution analyses of envelope standards and building form are provided in the appendices (see Appendices H and I).

Fig. 11 illustrates the urban-rural disparities in residential building energy codes in China. At the level of regulatory standards, energy conservation in residential buildings is first governed by national policies and laws and regulations, which provide the top-level framework for energy management. For instance, the Energy Conservation Law of the People's Republic of China serves as a fundamental legal document in the energy conservation domain. Similarly, the Action on Energy Efficiency for Civil Buildings was formulated to enhance energy efficiency management in civil buildings, reduce operational energy consumption, and improve overall energy utilization.

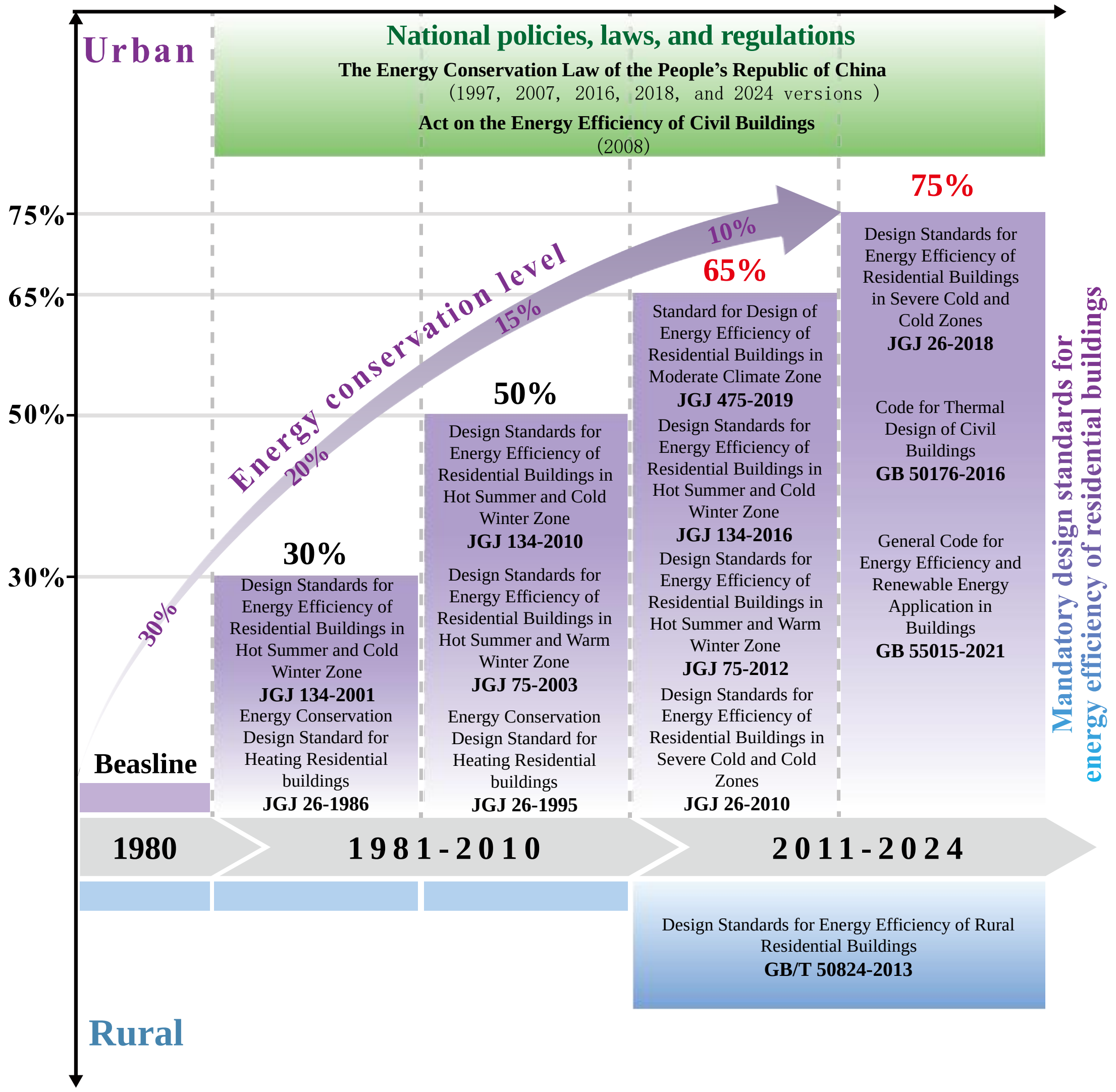

**Fig. 11.** Urban- rural disparities in residential building energy codes in China.

Under the guidance of these national policies and laws and regulations, a series of mandatory design standards for the energy conservation of residential buildings were developed. These standards regulate energy performance through indicators such as the thermal performance of the building envelope, the building shape coefficient, and the equipment energy efficiency ratio, thereby influencing the building's overall energy consumption. From 1980-2024, the mandatory design standards for energy conservation of residential buildings in urban areas were continuously updated and refined in accordance with China's five building thermal zones. In terms of improving energy efficiency, urban mandatory design standards for energy conservation of residential buildings began with the release of the energy conservation design standard for new heating residential buildings (JGJ 26-1986). These standards, which initially targeted SC and C zones, gradually expanded to include the M zone. Over time, the energy-saving targets have progressively increased from 30% to 50%, 65%, and ultimately 75% [48]. In contrast, during the study period, rural mandatory design standards for the energy conservation of residential buildings were largely limited in scope and authority. The only widely recognized and authoritative standard is the Design Standards for Energy Efficiency of Rural Residential Buildings (GB/T 50824-2013), released in 2013. Moreover, in terms of implementation, urban residential buildings are subject to strict approval, inspection, and regulatory control on the basis of established standards. In contrast, rural residential buildings are predominantly self-built, rely heavily on residents' voluntary adoption of energy-saving measures, and often lack systematic technical guidance and policy oversight. This disparity in administrative and enforcement mechanisms is a key factor contributing to the significant gap in energy conservation levels between urban and rural residential buildings.

Section 5.1 analyzes the evolving building stock patterns and the relationship between building prototypes and thermal loads in China's urban and rural areas while examining the underlying causes of urban-rural disparities, thereby resolving the second research question established in Section 1.

### *5.2. Policy recommendations*

Guided by the carbon neutrality strategy, the building sector, as a key area of China's energy consumption and carbon emissions, plays a decisive role in achieving the 2060 target [49, 50]. At present, China's building energy conservation efforts exhibit a distinct urban-rural divergence. Urban areas, supported by a well-established regulatory framework and mandatory standards, have made remarkable progress in improving building energy efficiency. In contrast, vast rural areas, due to the long-term absence of systematic policy support and technical guidance, have experienced severe delays in building energy conservation. This imbalance between urban and rural development not only constrains the overall emission reduction potential of the building sector but also falls short of the comprehensive requirements of the national carbon neutrality strategy. In particular, against the backdrop of intensifying global climate change, rural buildings face the dual challenge of addressing historical deficiencies while meeting the increasing demand for energy savings.

At present, China's urban residential buildings have established a relatively well-developed energy efficiency standard system, effectively reducing building energy consumption. In contrast, rural residential buildings still face significant shortcomings in terms of energy conservation. Owing to the absence of nationwide mandatory energy efficiency standards and effective regulatory measures, progress in rural building energy conservation has lagged behind. Existing regulations are mainly local and recommendatory in nature, and a unified national framework has yet to be formed. Notably, many rural buildings with heating loads, particularly in northern regions, which are dominated by space heating demand, have become an important component of decarbonization efforts in the building sector. At the same time, rural areas hold significant potential for the use of renewable energy, including dense biomass, solar photovoltaics, and solar heating [51]. The rational utilization of these resources could further promote energy conservation and carbon reduction in rural buildings [52]. From the perspective of regulatory frameworks, current rural energy efficiency standards suffer from two main limitations. On the one hand, documents such as the Technical Specification for Clean Heating in Rural Buildings [53] are purely technical guidance without binding force. On the other hand, although major energy-consuming provinces such as Hebei Province have issued local standards, including the Energy-Saving Design Standard for Rural Low-Energy Residential Buildings [54], regulatory gaps and enforcement disparities remain nationwide. In view of this situation, coordinated efforts are recommended in both institutional design and implementation supervision. At the standard level, a nationwide mandatory energy efficiency standard for rural residential buildings should be established, with a particular focus on incorporating envelope performance into the core indicator system [55], whereas northern provinces should be encouraged to benchmark against early adopters such as Hebei. At the implementation level, supporting measures such as improving the approval process for new buildings and establishing incentive mechanisms for the retrofitting of existing buildings should be introduced to ensure the effective enforcement of energy efficiency standards. Such systematic efforts will help narrow the urban-rural gap in building energy consumption and contribute to achieving the carbon neutrality goals.

Significant spatial heterogeneity exists in heating and cooling load characteristics across China's provinces. This variation arises not only from the climatic zoning induced by the country's wide latitudinal span but also from differences in regional economic development levels and the enforcement of building energy efficiency standards [56]. To narrow the interprovincial gap in building energy consumption, a regionally tiered control system should be established. Under the premise of maintaining unified national energy-saving targets, provinces should formulate differentiated implementation schemes on the basis of their local load characteristics. With global temperatures continuing to rise, the demand for cooling in buildings is increasing. To address this trend, climate-responsive and sustainable residential designs should be promoted, with a focus on the development of passive cooling technologies [57]. Measures such as building shading systems, optimized ventilation design, and improved thermal performance of building envelopes can be employed to effectively regulate indoor temperatures [3]. Within the structure of building energy consumption, heating loads remain dominant, particularly in SC and C zones. From 2000 to 2021, space heating made a significant contribution to the increase in carbon intensity, with China's residential heating growth being the highest in the world [58]. Given the substantial differences in climatic conditions, building types, and energy-use habits across regions, strategies for reducing heating loads must fully account for local characteristics and adopt context-specific technical pathways. Expanding electrification and achieving a cleaner power mix are essential for the transition to a low-carbon energy future [31]. Fuel switching, especially the transition from fossil fuels to clean electricity [59], can substantially reduce carbon emissions from building end uses [60, 61]. Therefore, the adoption of clean district heating measures [62], such as electric heat pumps [63] and waste heat recovery [64], is advocated [65]. For example, in provinces such as Liaoning and Heilongjiang, where heating demand is high, the transition from coal-based heating to clean energy should be prioritized. In provinces such as Henan, retrofitting existing buildings for energy conservation and upgrading heating systems can help reduce heating loads and improve efficiency [66]. Such regionally tailored measures can not only effectively narrow disparities in building energy consumption but also provide systematic solutions for carbon reduction in the national building sector. Section 5.2 extends the findings established in Section 5.1, with both sections collectively providing comprehensive resolution to the third research question identified in Section 1.

# 6. Conclusions

This study developed a bottom-up building energy model based on prototype buildings to assess heating and cooling loads in urban and rural residential buildings across 30 provinces in China from 1980 to 2024. The model quantified both total and per-floorspace heating and cooling loads at the provincial level, with particular attention to their spatial distribution and temporal evolution. It further examined the key factors driving urban-rural differences in residential thermal load patterns and proposed the corresponding policy recommendations. The core findings are summarized below.

## *6.1. Core findings*

- **The disparity in total heating and cooling loads between urban and rural residential buildings generally narrowed across most provinces over time.** For urban residential buildings, Guangdong consistently recorded the highest total cooling loads, increasing from 6.4 TWh/a in 1985 to 21.2 TWh/a in 2000, 42.5 TWh/a in 2010, and 76.5 TWh/a in 2020, representing the largest increase among all 30 provinces. In contrast, urban heating loads were highest in Liaoning in 1985, 2000, and 2010, with values of 11.7 TWh/a, 50.8 TWh/a, and 73.6 TWh/a, respectively, while Shandong ranked first in 2020, reaching 103.8 TWh/a. In rural residential buildings, Guangdong consistently exhibited the highest total cooling loads, rising from 34.0 TWh/a in 1985 to 45.9 TWh/a in 2000, 50.9 TWh/a in 2010, and 63.0 TWh/a in 2020. Rural heating loads were dominated by Henan throughout the study period, with values of 104.8 TWh/a in 1985, 201.6 TWh/a in 2000, 181.6 TWh/a in 2010, and 174.6 TWh/a in 2020, consistently exceeding 100 TWh/a. In contrast, Hainan remained the only province where rural heating loads were persistently below 1 TWh/a. Spatially, urban cooling load expanded from coastal provinces toward inland regions, while heating load remained concentrated in northern cold-climate areas. Although rural areas initially exhibited higher heating and cooling loads, urban areas in most provinces gradually caught up, with some cities—such as Beijing and Tianjin—surpassing rural areas in both heating and cooling loads after 2015.
- **Per-floorspace heating and cooling loads in rural residential buildings were generally higher than those in urban areas, particularly for heating in cold-climate regions.** From 1980 to 2024, the national average urban cooling load increased from 12.4 to 15.1 kWh/$m^2$·a, corresponding to an average annual growth rate of 0.44%. In contrast, rural cooling loads declined, with the national average decreasing from 22.63 to 19.87 kWh/$m^2$·a (-0.29% annually). Heating loads in both urban and rural areas exhibited downward trends: the national urban average fell from 44.08 to 39.92 kWh/$m^2$·a (-0.23% annually), while the rural average declined more sharply from 100.15 to 72.42 kWh/$m^2$·a (-0.73% annually). In provinces with relatively high cooling load, Hainan's urban cooling load increased from 31.09 kWh/$m^2$·a in 1980 to 33.94 kWh/$m^2$·a in 2024, whereas its rural cooling load decreased markedly from 82.36 to 47.14 kWh/$m^2$·a. A similar pattern was observed in Guangdong, where urban cooling loads rose from 27.73 to 29.71 kWh/$m^2$·a, while rural loads declined from 77.11 to 39.75 kWh/$m^2$·a. In provinces with high heating load, urban heating loads decreased from 132.20 to 119.59 kWh/$m^2$·a in Heilongjiang and from 122.47 to 111.85 kWh/$m^2$·a in Jilin. Rural heating loads in these provinces also declined substantially, from 339.91 to 197.58 kWh/$m^2$·a in Heilongjiang and from 291.92 to 176.11 kWh/$m^2$·a in Jilin, remaining the highest nationwide. Overall, urban areas experienced increasing cooling load intensities, whereas rural cooling loads declined. Heating loads decreased in both urban and rural areas, with the most pronounced reductions observed in northern regions. These results highlight strong climatic-zone dependence and persistent urban-rural disparities in per-floorspace heating and cooling load intensities.

- **The differences in heating and cooling loads between urban and rural residential buildings reflected the combined effects of building stock dynamics and envelope performance gaps.** From 2000 to 2024, substantial changes occurred in the distribution of urban and rural building stock, with the share of urban buildings surpassing that of rural buildings in 22 provinces, compared with only 4 provinces in 2000. This structural shift significantly reshaped the spatial patterns of heating and cooling loads across urban and rural areas. By 2024, heating and cooling loads in rural areas were generally 20%-100% higher than those in urban areas, primarily due to disparities in energy-efficiency regulation and implementation. Urban buildings have been governed by a rigorous and continuously updated system of mandatory energy-efficiency design standards, with twelve codes promulgated since the 1980s. In contrast, rural buildings have largely relied on a single standard—the Design Standards for Energy Efficiency of Rural Residential Buildings (GB/T 50824-2013)—issued in 2013. Owing to the absence of a comprehensive regulatory framework, limited technical guidance, and the prevalence of self-built construction, rural buildings often fail to fully implement prescribed standards, resulting in inferior envelope performance and persistently higher heating and cooling loads.

### *6.2. Forthcoming studies*

Several limitations of this study warrant further investigation. First, although this study provides a provincial-scale assessment of heating and cooling loads in China's urban and rural residential buildings from 1980 to 2024, the binary urban-rural classification remains relatively coarse. Introducing a hierarchical classification of urban settlements would help reveal differentiated trajectories and spatial patterns of thermal loads across diverse city types.

Second, building operational behavior is represented in a simplified manner. Key factors such as occupancy density, usage duration, and lifestyle patterns may substantially affect heating and cooling loads. To address this limitation, future research could leverage Internet of Things (IoT) technologies and cross-domain data-fusion techniques to incorporate more granular, real-time behavioral data into dynamic simulation frameworks, thereby improving model realism under varying socioeconomic and climatic conditions.

Furthermore, integrating heterogeneous data across 30 provinces, multiple climate zones, and distinct administrative categories requires improved system interoperability. Future studies could employ semantic representation, ontology-based data organization, and machine-learning-assisted feature extraction to enhance the scalability, reproducibility, and cross-domain data-integration capacity of multi-scale building-stock analyses. By linking dynamic building-performance data with broader socioeconomic indicators through interoperable frameworks, future research could establish a more rigorous quantitative basis for assessing policy equity related to persistent urban-rural gaps in residential heating and cooling loads.

Finally, the prototype-based bottom-up regional building-energy modeling framework developed in this study offers considerable flexibility and scalability and could be extended to other countries and climate zones to project future heating and cooling demand, thereby supporting building energy-efficiency improvements and climate-adaptation strategies.

## Acknowledgments

Q.T. appreciates the Fundamental Research Funds for the Central Universities of China (2026CDJSKCB04). N.Z. acknowledges support from the Energy Demand Changes Induced by Technological and Social Innovations network, an initiative coordinated by the Research Institute of Innovative Technology for the Earth and the International Institute for Applied Systems Analysis, and funded by the Ministry of Economy, Trade, and Industry, Japan.

## Data and Code Availability

The external data used in this study were obtained from statistical, regulatory, meteorological, and publicly accessible sources. Average residential floorspace per household in urban and rural areas was obtained from statistical yearbooks published by the National Bureau of Statistics of China and supplemented with province-specific statistics. Building geometry parameters were compiled from onsite surveys and publicly available platforms such as Zhulong.com, while envelope thermal-performance parameters were extracted from applicable building energy-efficiency design standards. Post-2000 meteorological data were obtained from Meteonorm 8 EnergyPlus Weather files, whereas pre-2000 meteorological data were obtained from the ECMWF Reanalysis v5 (ERA5) dataset. Internal heat-gain parameters were derived from published literature and measurements.

The author-compiled data include the prototype-building parameters and processed model inputs developed from these source datasets. Detailed data sources and processing procedures are described in Section 3.3, and the prototype-building parameters and urban-rural model settings are provided in Appendices B and C. The continuity assessment of the transition between the ERA5 and Meteonorm meteorological datasets is provided in Appendix G.

The model-derived data supporting the findings of this study and the EnergyPlus/OpenStudio prototype model files and aggregation scripts are available from the corresponding author upon reasonable request. Requests for additional information and resources should be directed to Dr. Minda Ma (minda.ma@cqu.edu.cn).

## Declaration of interests

The authors declare that they have no competing interests.